\documentclass[prd,twocolumn,aps,showpacs,nofootinbib,nobibnotes,superscriptaddress,preprintnumbers]{revtex4}

\usepackage{amsfonts,amssymb,amsmath}

\usepackage[figuresright]{rotating}
\usepackage{bm,color,amsmath,ulem}
\graphicspath{{./fig/}{./ps/}}
\preprint{NORDITA-2020-044, YITP-20-47, IPMU20-0042} 

\newcommand{\ypf}[4]{, ``#4,'' Phys.\ Fluids {\bf #2}, #3 (#1).}
\newcommand{\ybook}[3]{, {\it #2}. #3 (#1).}

\def\EEi{{\cal E}_i}
\def\EEM{{\cal E}_{\rm M}}
\def\xiM{\xi_{\rm M}}
\def\fourthird{{\textstyle\frac{4}{3}}}
\newcommand{\G}{\,{\rm G}}
\newcommand{\Mpc}{\,{\rm Mpc}}
\newcommand{\Fig}[1]{Fig.~\ref{#1}}
\newcommand{\Figs}[2]{Figs.~\ref{#1} and \ref{#2}}
\newcommand{\Tab}[1]{Table~\ref{#1}}
\newcommand{\bra}[1]{\langle #1\rangle}
\newcommand{\eee}{\hat{\mbox{\boldmath $e$}} {}}
\newcommand{\RRRR}{\mbox{\boldmath ${\sf R}$} {}}
\newcommand{\SSSS}{\mbox{\boldmath ${\sf S}$} {}}
\newcommand{\EEE}{\mbox{\boldmath ${\cal E}$} {}}
\newcommand{\FFF}{\mbox{\boldmath ${\cal F}$} {}}
\newcommand{\nab}{\bm{\nabla}}
\newcommand{\BB}{\mathbf{B}}
\newcommand{\ff}{\bm{f}}
\newcommand{\kk}{\bm{k}}
\newcommand{\uu}{\bm{u}}
\newcommand{\xx}{\bm{x}}
\def\fh{\tilde{{\bm{f}}}}
\def\EEK{{\cal E}_{\rm K}}
\def\EEM{{\cal E}_{\rm M}}
\def\HHM{{\cal H}_{\rm M}}

\def\EK{E_{\rm K}}
\def\EM{E_{\rm M}}
\def\cs{c_{\rm s}}

\definecolor{dgreen}{rgb}{0,0.7,0.0}
\newcommand{\rut}[1]{{\color{dgreen}#1}}

\begin{document}

\title{Primordial magnetic helicity evolution with a homogeneous magnetic field from inflation}

\author{Axel~Brandenburg}
\affiliation{Nordita, KTH Royal Institute of Technology and Stockholm University, Roslagstullsbacken 23, 10691 Stockholm, Sweden}
\affiliation{Department of Astronomy, AlbaNova University Center, Stockholm University, 10691 Stockholm, Sweden}
\affiliation{JILA and Laboratory for Atmospheric and Space Physics, University of Colorado, Boulder, Colorado 80303, USA}
\affiliation{McWilliams Center for Cosmology and Department of Physics, Carnegie Mellon University, 5000 Forbes Avenue, Pittsburgh, Pennsylvania 15213, USA}
\affiliation{Faculty of Natural Sciences and Medicine, Ilia State University, 3-5 Cholokashvili Street, 0194 Tbilisi, Georgia}

\author{Ruth~Durrer}
\affiliation{D\'epartement de Physique Th\'eorique and Center for Astroparticle Physics, Universit\'e de Geneve, Quai Ernest Ansermet 24, 1211 Gen\'eve 4, Switzerland}

\author{Yiwen~Huang}
\affiliation{McWilliams Center for Cosmology and Department of Physics, Carnegie Mellon University, 5000 Forbes Avenue, Pittsburgh, Pennsylvania 15213, USA}
\affiliation{Department of Physics, University of California San Diego, 9500 Gilman Drive 1110-115, La Jolla, California 92093}

\author{Tina~Kahniashvili}
\affiliation{McWilliams Center for Cosmology and Department of Physics, Carnegie Mellon University, 5000 Forbes Avenue, Pittsburgh, Pennsylvania 15213, USA}
\affiliation{Faculty of Natural Sciences and Medicine, Ilia State University, 3-5 Cholokashvili Street, 0194 Tbilisi, Georgia}
\affiliation{Abastumani Astrophysical Observatory, 47 Kostava Street, 0179 Tbilisi, Georgia}
\affiliation{Department of Physics, Laurentian University, Ramsey Lake Road, Sudbury, Ontario P3E 2C,Canada}

\author{Sayan~Mandal \footnote{Corresponding author; the authors are listed alphabetically.}}
\email{sayanm@andrew.cmu.edu}
\affiliation{McWilliams Center for Cosmology and Department of Physics, Carnegie Mellon University, 5000 Forbes Avenue, Pittsburgh, Pennsylvania 15213, USA}
\affiliation{Faculty of Natural Sciences and Medicine, Ilia State University, 3-5 Cholokashvili Street, 0194 Tbilisi, Georgia}

\author{Shinji~Mukohyama}
\affiliation{Center for Gravitational Physics, Yukawa Institute for Theoretical Physics, Kyoto University, Kyoto 606-8502, Japan}
\affiliation{Kavli Institute for the Physics and Mathematics of the Universe (WPI), The University of Tokyo Institutes for Advanced Study, The University of Tokyo, Kashiwa, Chiba 277-8583, Japan}

%\date{\today,~ $ $Revision: 1.125 $ $}

%%%%%%%%%%%%%%%%%%%%%%%%%%%%%%%%%%%%%%%%%%%%%%%%%%%%%%%
\begin{abstract}
Motivated by a scenario of magnetogenesis in which a homogeneous magnetic field is generated during inflation, we study the magnetohydrodynamic evolution of the primordial plasma motions
for two kinds of initial conditions --
(i) a spatially homogeneous field with an unlimited correlation length, and
(ii) a zero flux scale-invariant statistically homogeneous magnetic field.
In both cases, we apply, for a short initial time interval, monochromatic forcing
at a certain wave number so that the correlation length is finite, but much smaller
than the typical length scale of  turbulence.
In particular, we investigate the decay of nonhelical and helical hydromagnetic turbulence.
We show that, in the presence of a homogeneous magnetic field, the decay of
helical and nonhelical small-scale fields can occur rapidly.
This is a special property of a system with a perfectly homogeneous
magnetic field, which is sometimes considered as a local approximation
to a slowly varying background field.
It can never change and acts as an imposed magnetic field.
This is in a sharp contrast to the case of a statistically
homogeneous magnetic field, where we recover familiar decay properties: a much
slower decay of magnetic energy and a faster growth of the correlation length, especially
in the case with magnetic helicity.
The result suggests that a homogeneous magnetic field, if generated during inflation, should persist under the influence of small-scale fields and could be the origin of the large-scale magnetic field in the Universe. 
\end{abstract}
%%%%%%%%%%%%%%%%%%%%%%%%%%%%%%%%%%%%%%%%%%%%%%%%%%

\maketitle

\section{Introduction}
\label{intro}

One of several open problems in cosmology and astrophysics is the
understanding of the origin of large-scale magnetic fields
in the Universe \cite{Widrow:2002ud,Durrer:2013pga}.
There are two widely considered approaches to understand the origin of 
intercluster, large-scale correlated  magnetic fields --
(i) an \textit{astrophysical} scenario \cite{Kulsrud:2007an}, where weak seed fields generated by
local sources are amplified and transferred to large scales by various astrophysical processes, and
(ii) a \textit{cosmological} (or primordial) scenario \cite{Kandus:2010nw}, where a strong
seed magnetic field generated in the early universe evolves through magnetohydrodynamic (MHD)
coupling with the primordial plasma.
Neronov and Vovk \cite{Neronov:1900zz} used the nonobservation of GeV photons
from TeV blazars to put a
lower limit on the strength of magnetic fields on extragalactic scales,
obtaining a lower bound of $\sim 10^{-15}\,\mathrm{G}$ at $1\,\mathrm{Mpc}$.
The limits were later revised to $\sim 10^{-18}\,\mathrm{G}$ after
considering that the observation period of the sources was limited to
only a few years \cite{Dermer:2010mm,Taylor:2011bn}.
The Fermi-LAT and the VERITAS collaborations have improved this limit
again to $\sim 10^{-15}\,\mathrm{G}$ at $1\,\mathrm{Mpc}$
\cite{Biteau:2018tmv,Archambault:2017hvo}, based on ten years
of observations of the TeV blazar emission spectra.
These observational limits favor the cosmological (primordial) scenario of 
magnetogenesis \cite{Dolag:2010ni}; see Ref.~\cite{Arlen:2012iy}
for discussions on possible uncertainties in these lower limits based on blazar
spectra and Refs.~\cite{Broderick:2018nqf,AlvesBatista:2019ipr} on possible 
impacts of plasma instabilities.

There are several scenarios for generating primordial magnetic fields in the early universe;
see Ref.~\cite{Subramanian:2015lua} for a review.
Here we consider two main ideas.
First, magnetic fields can be generated during inflation or through processes related to it,
like \textit{reheating} or \textit{preheating}.
Reheating is the epoch at the end of inflation when the energy in
the hypothesized inflaton field decays into the fields of the standard model and
the temperature of the universe rises sufficiently.
The decay of the inflaton into bosons can be very rapid owing to processes 
such as a parametric resonance or a tachyonic instability. 
Such a rapid decay is called preheating.
Primordial magnetic fields can also be generated during cosmological phase transitions.
The evolution of these primordial fields in the expanding universe has been
studied by several authors by solving the 
MHD equations for the magnetic field,
the density, and the velocity of the plasma;
see Ref.~\cite{Kahniashvili:2018mzl} for a brief review and references within.
The fields used are generally
modeled either as {\it homogeneous}, or as {\it statistically homogeneous}
and isotropic random Gaussian stochastic fields.
Statistical homogeneity implies that the two-point correlation function
of the magnetic field is independent of the position in space.
In this paper, we show that these two approaches can result
in very different
dynamics of the induced turbulent motions in the early universe.
In particular for the former case, small-scale helical and nonhelical fields
decay in a way very different from the case of statistically homogeneous fields,
as discussed below.

While there were earlier ideas suggesting the presence of a homogeneous
magnetic field~\cite{Zeldovich:1965,Doroshkevich:1965,Thorne:1967zz},
those papers study just cosmological consequences without specifying
or discussing a generation mechanism. While Ref.~\cite{Harrison:1969}
discusses the generation of a homogeneous magnetic field,
it assumes the existence of ``protogalaxies'' with angular momentum
in the radiation dominated epoch, which contradicts with the
current understanding of our Universe. Reference~\cite{Bertolami:1998dn}
also discusses a generation mechanism but it assumes a tachyonic mass
for a gauge field, which means that the corresponding Higgs field
is a ghost and thus the model is unstable. While the model in
Ref.~\cite{Maeda:2012eg} is similar to gauge-flation or
chromo-natural inflation, what is called a magnetic field
there is simply due to the nonlinear part of the field strength
and is thus actually an electric field, if a part of the
non-Abelian gauge field is projected onto an Abelian gauge field
via a spontaneous symmetry breaking (e.g., ${\rm SU}(2)\times {\rm U}(1) \to {\rm U}(1)$)
or if their ansatz of the gauge potential is applied to an
Abelian gauge field.
As far as the authors know, no stable generation mechanism
of homogeneous magnetic fields has been proposed until recently.
(In open Universes, while one might hope to find such a mechanism
through supercurvature modes, but it is known that there is actually
no supercurvature mode for vector fields [14].)
However, the lack of a generation mechanism does not necessarily
mean a lack of interest in homogeneous magnetic fields.
Indeed, as already mentioned above, the studies of cosmological
consequences of a homogeneous magnetic field date back to the
seminal works by Zel'dovich, Doroshkevich, and Thorne in 1960'=s.

Recently, one of the authors \cite{Mukohyama:2016npi} proposed
a stable generation mechanism of homogeneous magnetic field, based
on a $U(1)$ gauge theory of electromagnetism with a coupling to
Horndeski type \textit{scalar-tensor gravity}, where gravity is
described by the metric tensor field and an additional scalar
field\footnote{See, e.g., \cite{Fujii:2003pa} for more details
about scalar-tensor gravity.}.
During inflation, the model admits a stable (quasi) de Sitter solution
with a homogeneous magnetic field as an attractor of the system.
Therefore the model provides a classical generation mechanism
of homogeneous magnetic fields during inflation.
After inflation and the stabilization of the scalar field at
the minimum of its potential, on the other hand, gravity is
effectively described by general relativity.
Later in the present paper, we show that the action after inflation is the Einstein-Maxwell action,
supplemented with a nonminimal coupling between curvature and electromagnetism.
We also show that, upon imposing observational constraints, the nonminimal 
coupling can be ignored for the analysis of the magnetic field evolution.
This in particular means that the no-minimal coupling does not introduce
new instabilities in the homogeneous magnetic field background in the
late-time cosmology. 

It was already known that in the additional presence of primordial
small-scale turbulence, the magnetic energy spectrum changes only very
little at large length scales \cite{Kahniashvili:2012vt}.
This led one of the authors \cite{Mukohyama:2016npi} to expect that this also
applies to the case of a homogeneous magnetic field, but this remained
to be verified by numerical MHD simulations.
It is therefore important to study the MHD evolution of primordial plasma
motions in the presence of these homogeneous magnetic fields, which is what we
focus on in this work.

We are  particularly interested in the evolution of magnetic helicity,
which is known not to be conserved in a periodic domain in the presence of
a homogeneous magnetic field \cite{Ber97}.
However, if we were to consider a perfectly homogeneous magnetic field
as a local approximation to a slowly varying background magnetic field,
magnetic helicity conservation would be restored.
To illuminate the remarkable properties of a perfectly homogeneous
magnetic field, we also discuss the alternative approach of working instead
with a statistically homogeneous magnetic field, which does not
impose any constraints on the magnetic helicity evolution.
The presence of magnetic helicity substantially changes the decay rate
for MHD turbulence \cite{BM99}.
In this paper we compare the decay dynamics for homogeneous and
statistically homogeneous magnetic fields with a scale-invariant spectrum.
In previous works, we have studied only statistically homogeneous
magnetic fields induced by the turbulence dynamics \cite{Brandenburg:2016odr} 
but did not include the turbulence in the presence of a homogeneous magnetic field.

This paper is arranged as follows.
The model is described in Sec.~\ref{secModel},
where we discuss the formalism for how a spatially homogeneous magnetic
field is realized during inflation, and after that until recombination.
In Sec.~\ref{numEvol}, we describe in detail the setup of our simulations,
discussing, in particular, various initial conditions to examine
peculiar features associated with the use of an imposed magnetic field.
We present numerical solutions in Sec.~\ref{secResults}
and in Sec.~\ref{secConcl}, we present our conclusions.
Throughout this paper we work in natural units where $\hbar=c=1$,
and our metric signature is $(-,+,+,+)$. For the electromagnetic quantities we use Lorentz-Heaviside units.

\section{Homogeneous magnetic fields}
\label{secModel}

In this section we briefly describe a theoretical framework in which a spatially
homogeneous magnetic field background can be realized during and after inflation
in the early Universe.
In the inflationary stage, the background spacetime is not only homogeneous but
also isotropic despite the existence of the preferred spatial direction defined
by the homogeneous magnetic field.
This is made possible by a nonlinear kinetic action for the $U(1)$ gauge field
nonminimally coupled to a scalar-tensor theory of gravity.
In the postinflationary stage, on the other hand, the scalar field is stabilized
around a minimum of a potential and thus the theory is reduced to the
Einstein-Maxwell theory supplemented with the Horndeski's nonminimal coupling.
Therefore, after inflation the spacetime becomes anisotropic and the homogeneous
magnetic field adiabatically decays.
If we are interested in the postinflationary evolution of the $U(1)$ gauge field at
subhorizon scales for timescales sufficiently shorter than the cosmological time
then the gravitational effects of and on the gauge field can be neglected and the
system is described by the standard Maxwell theory expanded around the homogeneous
magnetic field background in Minkowski spacetime.
As we shall see in the next sections, the existence of the homogeneous magnetic
field significantly affects the evolution of the gauge field at subhorizon scales.

\subsection{General action}
\label{subsec:generalaction}

We consider a metric $g_{\mu\nu}$, a $U(1)$ gauge field $A_{\mu}$ and a scalar
field $\phi$ in four dimensional spacetime described by the action
\begin{equation}
 I = \int d^4x \sqrt{-g}
  \left[ L + L_3 + L_4 + L_5 + L_{\rm H}\right]\,, \label{eqn:action}
\end{equation}
where $L=L(\phi, X,W,Y,Z)$ is an arbitrary function of $\phi$, 
\begin{align}
 &
  X \equiv  -\frac{1}{2}g^{\mu\nu}\partial_{\mu}\phi\partial_{\nu}\phi\,,\quad
  W \equiv  -\frac{1}{4}\mathcal{F}_{\mu\nu}\mathcal{F}^{\mu\nu}\,,\nonumber\\
 &
  Y \equiv  \mathcal{F}_{\mu\nu}\tilde{\mathcal{F}}^{\mu\nu}\,,\quad
 Z \equiv  \mathcal{F}^{\rho\mu}\mathcal{F}_{\rho}^{\ \nu}\partial_{\mu}\phi\partial_{\nu}\phi\,; 
\end{align}
$\mathcal{F}_{\mu\nu}$ and $\tilde{\mathcal{F}}^{\mu\nu}$ are defined by 
\begin{align}
& \mathcal{F}_{\mu\nu} \equiv e^{\phi}F_{\mu\nu}\,,
  \quad
  \tilde{\mathcal{F}}^{\mu\nu} \equiv e^{\phi}\tilde{F}^{\mu\nu}\,, \nonumber\\
& 
 F_{\mu\nu} \equiv \partial_{\mu}A_{\nu} - \partial_{\nu}A_{\mu}\,, \quad
  \tilde{F}^{\mu\nu} \equiv \frac{1}{2}\epsilon^{\mu\nu\rho\sigma}F_{\rho\sigma}\,,
\end{align}
and $\epsilon^{0123}=-1/\sqrt{-g}$; 
\begin{align}
 L_3 =& -G_3(\phi,X)\Box\phi\,, \nonumber\\
 L_4 =& G_4(\phi,X) R + G_{4X}(\phi,X)\left[(\Box\phi)^2-(\nabla^{\mu}\nabla_{\nu}\phi)(\nabla^{\nu}\nabla_{\mu}\phi)\right]\,,\nonumber\\
 L_5 =& G_5(\phi,X)G^{\mu\nu}\nabla_{\mu}\nabla_{\nu}\phi 
  - \frac{1}{6}G_{5X}(\phi,X) 
  \left[ (\Box\phi)^3 \right.\nonumber\\
&  - 3(\Box\phi)(\nabla^{\mu}\nabla_{\nu}\phi)(\nabla^{\nu}\nabla_{\mu}\phi) \nonumber\\
& \left.+ 2(\nabla^{\mu}\nabla_{\nu}\phi)(\nabla^{\nu}\nabla_{\rho}\phi)(\nabla^{\rho}\nabla_{\mu}\phi) \right]
\end{align}
are Horndeski scalar terms~\cite{Horndeski:1974wa,Deffayet:2011gz}; and
\begin{equation}
 L_{\rm H} = \xi(\phi)\tilde{\mathcal{F}}^{\mu\nu}\tilde{\mathcal{F}}^{\rho\sigma}R_{\mu\nu\rho\sigma}
\end{equation}
is a simple modification of Horndeski's nonminimal coupling of the $U(1)$
gauge field to the Riemann tensor $R^{\mu}_{\ \nu\rho\sigma}$ of the metric
$g_{\mu\nu}$~\cite{Horndeski:1976gi}.
Here, the scalar field $\phi$ and the gauge field $A_{\mu}$ are normalized so
that their mass dimensions are zero, $G_{3,4,5}(\phi, X)$ are arbitrary
functions of $\phi$ and $X$, the subscript $X$ denotes derivative with respect
to $X$, and $\xi(\phi)$ is an arbitrary function of $\phi$.
The action is invariant under the $U(1)$ gauge transformation,
\begin{equation}
 A_{\mu} \to A_{\mu} + \partial_{\mu} \lambda\,,   \label{eqn:U(1)gaugesymmetry}
\end{equation}
where $\lambda$ is an arbitrary function, and the equations of motion are
second-order differential equations.
In principle it is possible to consider a more general form of $L$ that depends
on the second covariant derivatives of $\phi$ and $A_{\mu}$ without introducing
higher derivatives in the equations of motion.
For simplicity, however, we restrict our consideration to the above form of $L$
that depends on only up to first derivatives of $\phi$ and $A_{\mu}$.
Also, the inclusion of the factor $e^{\phi}$ in the definitions of
$\mathcal{F}_{\mu\nu}$ and $\tilde{\mathcal{F}}^{\mu\nu}$ is redundant since
we allow for the explicit $\phi$-dependence of $L(\phi, X,W,Y,Z)$ and $\xi(\phi)$.
We nonetheless adopt the above definitions of $\mathcal{F}_{\mu\nu}$ and
$\tilde{\mathcal{F}}^{\mu\nu}$ including the factor $e^{\phi}$ in order to make
it easy to implement a scaling-type symmetry for the description of the system
during the inflationary stage [see Eqs.~(\ref{eqn:scaling-symmetry}]
and (\ref{eqn:G345-xi-nophi}) in the next subsection). 

\subsection{Stealth magnetic field during inflation}

Following the discussion in Sec. V of \cite{Mukohyama:2016npi}, we suppose
that the main source of curvature perturbations is not $\phi$ but something else.
For example, one can introduce another scalar field as an inflaton or a curvaton.
For simplicity we approximate the geometry during inflation by a de Sitter spacetime.
Then the effective cosmological constant induced by the field responsible for
curvature perturbations simply amounts to a constant shift of $L(\phi, X,W,Y,Z)$.

In order to simplify the analysis and also to allow for an exact solution that
represents a de Sitter spacetime with a homogeneous magnetic field, we require
that the action is invariant under not only the $U(1)$ gauge transformation
(\ref{eqn:U(1)gaugesymmetry}) but also the following scaling-type global
transformation for the range of $\phi$ that is relevant for the inflationary epoch.
\begin{equation}
 \phi \to \phi + \phi_0\,, \quad A_{\mu} \to e^{-\phi_0}A_{\mu}\,,\label{eqn:scaling-symmetry}
\end{equation}
where $\phi_0$ is an arbitrary constant that is not too large to eject $\phi$
from the inflationary range.
Then for the range of $\phi$, the explicit $\phi$-dependence of the functions
$L(\phi,X,W,Y,Z)$, $G_{3,4,5}(\phi, X)$ and $\xi(\phi)$ is forbidden so that
\begin{eqnarray}
 L(\phi,X,W,Y,Z) &= &\bar{L}(X,W,Y,Z)\,, \nonumber\\
G_{3,4,5}(\phi,X) & = & \bar{G}_{3,4,5}(X)\,, \nonumber \\
\xi(\phi) & = & \bar{\xi}\,, \label{eqn:G345-xi-nophi}
\end{eqnarray}
where $\bar{L}(X, W, Y, Z)$ is an arbitrary function of ($X$, $W$, $Y$, $Z$), $\bar{G}_{3,4,5}(X)$
are arbitrary functions of $X$ and $\bar{\xi}$ here is a constant.
We also impose the parity invariance so that the function $\bar{L}(X,W,Y,Z)$ is
even with respect to $Y$. 
\begin{equation}
\bar{L}(X,W,Y,Z) = \bar{L}(X,W,-Y,Z)\,. \label{eqn:parityinvariance}
\end{equation}
This is the system studied in \cite{Mukohyama:2016npi,Mukohyama:2018obj}. 

For this system, we adopt the ansatz of the form
\begin{eqnarray}
&& g_{\mu\nu}dx^{\mu}dx^{\nu} = 
 -N(t)^2dt^2 \nonumber\\
& & \qquad\qquad + a(t)^2 \left[ e^{4\sigma(t)}dx^2 + e^{-2\sigma(t)}(dy^2+dz^2)\right]\,,\nonumber\\
&& \phi = \phi(t)\,, \nonumber\\
&& A_t = 0\,,\quad A_x = \int^t \frac{N(t')e^{4\sigma(t')}}{a(t')}E(t')dt'\,, \nonumber\\
&& A_y = \frac{1}{2}Bz\,,\quad   A_z = -\frac{1}{2}By\,,  \label{eqn:ansatz}
\end{eqnarray}
where $B$ is a constant. It was found in \cite{Mukohyama:2016npi} that the
equations of motion admit solutions of the form
\begin{eqnarray}
&& H = \mathrm{const.} > 0\,, \quad \Sigma = \mathrm{const.}\,, \quad \chi = \mathrm{const.} > 0\,, \nonumber\\
&& E = \mathrm{const.}\,, \quad B\ne 0\,, 
\end{eqnarray}
where
\begin{equation}
 H \equiv \frac{\dot{a}}{Na}\,, \quad 
  \Sigma \equiv \frac{\dot{\sigma}}{N}\,,\quad
  \chi \equiv \frac{e^{\phi}e^{2\sigma}}{a^2}\,.
\end{equation}
By tuning one parameter in the action, the solution is reduced to a de Sitter
spacetime with magnetic field but without electric field~\cite{Mukohyama:2016npi}, i.e. 
\begin{eqnarray}
&& H = \mathrm{const.} > 0\,, \quad \Sigma = 0\,, \quad
\chi = \mathrm{const.} > 0\,, \nonumber\\
&& E = 0\,, \quad B\ne 0\,.
\end{eqnarray}
The reason why fine-tuning of just one parameter leads to two equalities, $\Sigma=0$
and $E=0$, is that we have imposed the discrete symmetry (\ref{eqn:parityinvariance}).
Reference~\cite{Mukohyama:2016npi} also found the condition under which the de Sitter
solution with magnetic field but without electric field is an attractor of the
system within the ansatz (\ref{eqn:ansatz}).
Reference~\cite{Mukohyama:2018obj} then analyzed general linear perturbations around
the attractor solution and found the condition under which the system of linear
perturbations is free from instabilities. 

In the present paper we consider the stable attractor de Sitter solution with
magnetic field but without electric field as the origin of magnetic fields that
are observed in the late-time Universe.
We denote the (approximately) constant value of $H$ during inflation as
$H_{\rm inf}$~\footnote{In Refs.~\cite{Mukohyama:2016npi,Mukohyama:2018obj} it
was denoted as $H_0$.}. 

\subsection{Postinflationary system}

Following again the discussion in Sec. V of \cite{Mukohyama:2016npi},
we suppose that the scaling-type global symmetry (\ref{eqn:scaling-symmetry})
is not respected for the range of $\phi$ that is relevant for the postinflationary
epoch so that the scalar field $\phi$ is stabilized at a local minimum of a potential,
which we denote as $\phi_f$.
The action of the system is still supposed to be of the general form considered
in Sec. \ref{subsec:generalaction}.
Assuming that the mass of $\phi$ around the local minimum of the potential is
large enough, we integrate out $\phi$ by setting $\phi = \phi_f$
(and thus $X = 0$ and $\nabla_{\mu}\nabla_{\nu}\phi = 0$) in the general action.
We then end up with the following action for the system after inflation. 
\begin{eqnarray}
 I & = & \int d^4x\sqrt{-g} \left[ G_4(\phi_f,0) R + L(\phi_f, 0, W_f, Y_f, 0)\right.\nonumber\\
& & \left. + \xi(\phi_f) e^{2\phi_f}F_{\mu\nu}F_{\rho\sigma}R^{\mu\nu\rho\sigma} \right]\,,
\end{eqnarray}
where
\begin{equation}
  W_f \equiv -\frac{1}{4}e^{2\phi_f}F_{\mu\nu}F^{\mu\nu}\,, \quad
   Y_f \equiv e^{2\phi_f}F_{\mu\nu}\tilde{F}^{\mu\nu}\,. 
\end{equation}
By Taylor expanding $L(0, W_f, Y_f, 0)$ with respect to $W_f$ and $Y_f$ up to
first order and using the discrete symmetry (\ref{eqn:parityinvariance}),
we obtain the low-energy effective action 
\begin{eqnarray}
 I & = & \int d^4x\sqrt{-g} \left[ \frac{M_{\rm Pl}^2}{2} (R-2\Lambda) - \frac{1}{4} F^{\rm (post)}_{\mu\nu}F^{{\rm (post)}\mu\nu}\right.\nonumber\\
& & \left. + \frac{\lambda}{4M_{\rm Pl}^2}\tilde{F}^{\rm (post)}_{\mu\nu}\tilde{F}^{\rm (post)}_{\rho\sigma}R^{\mu\nu\rho\sigma} \right]\,, \label{eqn:action-after-inflation}
\end{eqnarray}
where we have assumed that
\begin{equation}
  G_4(\phi_f, 0) > 0\,, \quad L_W(\phi_f, 0,0,0,0) > 0\,,
\end{equation}
and introduced
\begin{equation}
 M_{\rm Pl} \equiv \sqrt{2 G_4(\phi_f, 0)}\,, \quad \Lambda \equiv - \frac{L(\phi_f, 0,0,0,0)}{M_{\rm Pl}^2}\,,
\end{equation}
\begin{equation}
 \lambda \equiv \frac{4M_{\rm Pl}^2\xi(\phi_f)}{L_W(\phi_f, 0,0,0,0)}\,, \ 
  F^{\rm (post)}_{\mu\nu} = e^{\phi_f} \sqrt{L_W(\phi_f, 0,0,0,0)} F_{\mu\nu}\,,\nonumber
\end{equation}
and 
\begin{equation}
 \tilde{F}^{\rm (post)}_{\mu\nu} \equiv \frac{1}{2}{\epsilon_{\mu\nu}}^{\rho\sigma} F^{\rm (post)}_{\rho\sigma}\,.
\end{equation}
Here, the subscript $W$ denotes partial derivative with respect to \ $W$. 
So far, we have not yet fixed the overall normalization of $F_{\mu\nu}$ except
that the mass dimension of $A_{\mu}$ is zero.
We now fix the normalization as
\begin{equation}
 e^{2\phi_f} L_W(\phi_f, 0,0,0,0) = M_{\rm Pl}^2\,,
\end{equation}
so that
\begin{equation}
\lambda \equiv 4e^{2\phi_f}\xi(\phi_f)\,, \quad 
 F^{\rm (post)}_{\mu\nu} = M_{\rm Pl}F_{\mu\nu}\,. \label{eqn:def-lambda-Fpost}
\end{equation}
The postinflationary system described by the action (\ref{eqn:action-after-inflation})
is nothing but the Einstein-Maxwell system supplemented with the Horndeski's nonminimal coupling.

Hereafter, we omit the superscript ``${\rm (post)}$'' so that the action for the
postinflationary system is
\begin{eqnarray}
 I & = & \int d^4x\sqrt{-g} \left[ \frac{M_{\rm Pl}^2}{2} (R-2\Lambda) - \frac{1}{4} F_{\mu\nu}F^{\mu\nu}\right.\nonumber\\
& & \left. + \frac{\lambda}{4M_{\rm Pl}^2}\tilde{F}_{\mu\nu}\tilde{F}_{\rho\sigma}R^{\mu\nu\rho\sigma} \right]\,.
\end{eqnarray}

\subsection{Observational bounds on $\lambda$, $H_{\rm inf}$, and $\sigma$}

In Ref.~\cite{Mukohyama:2016npi}, assuming that the stabilization of $\phi$ to
the constant value $\phi_f$ occurs immediately after inflation and that the
reheating process is instantaneous, the present amplitude of the large-scale
magnetic field was estimated as
\begin{equation}
\mathcal{B}_{\rm today} \simeq e^{-\phi_f}|b| \times 10^{-6}\,\mathrm{G}\,,
\label{eqn:Btoday}
\end{equation}
where $b \equiv B/H_{\rm inf}$.
Also, \cite{Mukohyama:2018obj} found several examples of parameters for which
the system of linear perturbations is free from instabilities.
In those examples, both $b$ and $g_h$ are nonvanishing and of order unity, where
\begin{equation}
 g_h \equiv \xi \frac{H_{\rm inf}^2}{M_{\rm Pl}^2}\,,
\end{equation}
and $\xi$ is the constant value of $\xi(\phi)$ for the range of $\phi$ relevant
for the inflationary epoch as already stated around (\ref{eqn:G345-xi-nophi}).
Under the assumption of immediate stabilization of $\phi$ after inflation, we
have $\xi(\phi_f) = \xi$.
Combining all these and the definition of $\lambda$ given in (\ref{eqn:def-lambda-Fpost}),
one obtains
\begin{equation}
\lambda \simeq 4\times \left(\frac{\mathcal{B}_{\rm today}}
{|b|\times 10^{-6}\,\mathrm{G}}\right)^{-2}
\left(\frac{H_{\rm inf}}{M_{\rm Pl}}\right)^{-2}g_h\,. \label{eqn:lambda-Hinf}
\end{equation}

The upper bound on the large scale magnetic field is roughly
$10^{-9}\,\mathrm{G}$~\cite{Shaw:2010ea} and the lower bound from the blazar
observations is roughly
\begin{equation}
10^{-15}\,\mathrm{G} \lesssim \mathcal{B}_{\rm today} \lesssim 10^{-9}\,\mathrm{G}\,. \label{eqn:Btoday-bound}
\end{equation}
On the other hand, constraints on $\lambda$ can be obtained by demanding
that the nonminimal coupling term is less important than the standard
Maxwell term~\cite{Barrow:2012ay}.
Reference~\cite{Allahyari:2020jkn} applied this idea to neutron stars and
found a conservative bound on $\lambda$ as
\begin{equation}
|\lambda| \ll 10^{70}\,. \label{eqn:lambda-bound}
\end{equation}

Combining (\ref{eqn:lambda-bound}) with (\ref{eqn:lambda-Hinf}), one obtains
a lower bound on the inflation scale,
\begin{equation}
H_{\rm inf} \gg |b|\, |g_h|^{1/2}\,
\left(\frac{\mathcal{B}_{\rm today}}{10^{-9}\,\mathrm{G}}\right)^{-1}
\times 10^{-15}\,\mathrm{GeV}\,.
\label{eqn:Hinf-bound}
\end{equation}
For the range (\ref{eqn:Btoday-bound}) of $\mathcal{B}_{\rm today}$ and
$\mathcal{O}(1)$ values of $b$ and $g_h$, this is not a strong constraint.
Under the assumption of instantaneous reheating ($T_{\rm reh}\sim \sqrt{M_{\rm Pl}H_{\rm inf}}$),
Eq.~(\ref{eqn:Hinf-bound}) can be rewritten as a lower bound on the reheating temperature.
\begin{equation}
T_{\rm reh} \gg |b|^{1/2}\, |g_h|^{1/4}\, \left(\frac{\mathcal{B}_{\rm today}}{10^{-9}\,\mathrm{G}}\right)^{-1/2} \times 100\,\mathrm{GeV}\,.
\end{equation}

One can also obtain limits on the parameter $\sigma$ which characterizes the
degree of axisymmetry of the Bianchi-I spacetime from its contribution to the
quadrupole component $C_2$ of the power spectrum of temperature anisotropies of
the cosmic microwave background (CMB).
This contribution can be written as $C_2=16\pi(\sigma_{\rm dec}-\sigma_{0})^2/25$
(see Ref.~\cite{Adamek:2011pr} for an outline of the calculation), where
$\sigma_{\rm dec}$ and $\sigma_{0}$ are values of $\sigma$ at the decoupling
and at the present, respectively. 
From the observed CMB quadrupole of $C_2^{\rm obs}=230\,\mu\mathrm{K}^2/T_0^2$,
where $T_0$ is the CMB temperature today.
We can always normalize our coordinates such that $\sigma_0=\sigma(t_0)=0$
so that $C_2$ provides an upper bound on $|\sigma_{\rm dec}|$,
\begin{equation}\label{eSigUpLim}
|\sigma_{\rm dec}| \lesssim 4\times 10^{-6}\,.
\end{equation}

\subsection{Subhorizon description of postinflationary system}

In general the effects of the nonminimal coupling can be ignored if 
\begin{equation}
 \frac{(\mbox{curvature})}{M_{\rm Pl}^2} \ll \frac{1}{|\lambda|}\,. \label{eqn:negligible-non-minimal-coupling}
\end{equation}
For the Friedmann-Lema\^{i}tre-Robertson-Walker (FLRW) cosmology, we have $(\mbox{curvature}) \sim T^4/M_{\rm Pl}^2$,
where $T$ is the temperature of the Universe, and thus the nonminimal coupling
can be ignored if 
\begin{equation}\label{eqn:tem-nonmim-coup-ignore}
T \ll \left|\frac{\lambda}{10^{70}}\right|^{-1/4}\times 10\,\mathrm{GeV}\,.
\end{equation}
Therefore, imposing the conservative bound (\ref{eqn:lambda-bound}),
we conclude that the evolution of the FLRW background cosmology
during and after nucleosynthesis
can be described by the standard Einstein-Maxwell theory without the nonminimal coupling.
For a local magnetic field with the amplitude $\mathcal{B}_{\rm local}$, the
induced curvature is of order $(\mbox{curvature}) \sim \mathcal{B}_{\rm local}^2/(8\pi M_{\rm Pl}^2)$
and thus the nonminimal coupling can be ignored if
\begin{equation}
\mathcal{B}_{\rm local} \ll \left|\frac{\lambda}{10^{70}}\right|^{-1/2}\times 10^{21}\,\mathrm{G}\,.
\end{equation}
Assuming that the conservative bound (\ref{eqn:lambda-bound}) is satisfied, the
right-hand side is larger than $10^{21}\,\mathrm{G}$ and thus the maximum
amplitude of the magnetic field in 
the simulations studied in the next sections satisfies this condition.
Therefore, we can safely ignore the effects of the nonminimal coupling and the
theory is reduced to the standard Einstein-Maxwell theory without the nonminimal coupling.

For the standard Einstein-Maxwell theory in a radiation dominated Universe
without the nonminimal coupling,
the propagation speed of all physical degrees of freedom is of order unity
and the Jeans scale is of order the Hubble scale.
If we are interested in phenomena whose length and timescales are sufficiently
shorter than the Jeans scales and the cosmological scales then the evolution of
the system can be well described without taking into account the metric
perturbation and the background cosmological expansion.
On these scales, the system is well described by the standard Maxwell theory
expanded around the homogeneous magnetic field background in Minkowski spacetime.

\section{Magnetic field evolution}
\label{numEvol}

In the previous section, we have discussed how a spatially homogeneous magnetic
field can be realized during inflation, and more importantly, after the end
of inflation.
We now turn our attention to the study of the MHD evolution
of such fields.

\subsection{Basic equations}

We study the time evolution in the presence of a homogeneous magnetic field
right after inflation.
In particular, we study the evolution of an additional field
with some typical wave number $k_*$, which we induce by a random
forcing term present during a short initial time interval.
In the radiation dominated era, the primordial plasma is a relativistic,
isothermal gas with energy density $\rho$ and equation of state $w=1/3$.
In Lorentz-Heaviside units, the MHD equations for such a gas
are \cite{Brandenburg:1996fc,Kahniashvili:2016bkp,Brandenburg:2017neh}
\begin{eqnarray}
\label{dlnrhodt}
\frac{\partial\ln\rho}{\partial t}&=&
-\frac{4}{3}\left(\nabla\cdot\mathbf{u}
+\mathbf{u}\cdot\nabla\ln\rho\right) \nonumber \\
&&+\frac{1}{\rho}\left[\mathbf{u}\cdot
(\mathbf{J}\times\mathbf{B})+\eta\mathbf{J}^2\right], \\
\frac{\partial\mathbf{u}}{\partial t}&=&
-(\mathbf{u}\cdot\nabla)\mathbf{u}
+\frac{\mathbf{u}}{3}\left(\nabla\cdot\mathbf{u}
+\mathbf{u}\cdot\nabla\ln\rho\right) \nonumber \\
&&-\frac{1}{4}\nabla\ln\rho
+\frac{3}{4\rho}\mathbf{J}\times\mathbf{B}
+\frac{2}{\rho}\nabla\cdot\left(\rho\nu\SSSS\right) \nonumber \\
&&-\frac{\mathbf{u}}{\rho}\left[\mathbf{u}\cdot
(\mathbf{J}\times\mathbf{B})+\eta\mathbf{J}^2\right]
+\FFF_0, \\
\frac{\partial\mathbf{B}}{\partial t}&=&
\nabla\times(\mathbf{u}\times\mathbf{B}-\eta\mathbf{J})+\EEE_0,
\label{dAdt}
\end{eqnarray}
where $\mathsf{S}_{ij}=\frac{1}{2}(u_{i,j}+u_{j,i})-\frac{1}{3}\delta_{ij}\nabla\cdot\mathbf{u}$
are the components of the traceless rate-of-strain tensor, $\nu$ is the kinematic viscosity,
$\eta$ is the magnetic diffusivity,
$\FFF_0={\cal F}_0\ff$ and $\EEE_0={\cal E}_0\ff$ are forcing terms, and
\begin{equation}
\ff(\xx,t)={\rm Re}\{{\cal N}\fh(\kk,t)\exp[i\kk\cdot\xx+i\phi]\},
\end{equation}
is a forcing function that consists of random, white-in-time,
plane waves with a certain average wave number $k_*$ \cite{Brandenburg:2000wm}.
Here, $\xx$ is the position vector and ${\cal N}=\sqrt{\cs^3 k_*}$
is a normalization factor with $\cs=\sqrt{w}=1/\sqrt{3}$ being the
speed of sound; see Ref.~\cite{Brandenburg:2000wm} for details.
At each time step, we select randomly the phase $-\pi<\phi\le\pi$,
the direction of a unit vector $\eee$, and the components of the
wave vector $\kk$ from many possible discrete wave vectors in a certain
range around a given value of $k_*$.
The Fourier amplitudes are
\begin{equation}
\fh({\kk})=\RRRR\cdot\fh({\kk})^{\rm(nohel)}\quad\mbox{with}\;\;
{\sf R}_{ij}=\frac{\delta_{ij}-i\sigma\epsilon_{ijk}\hat{k}}
{\sqrt{1+\sigma^2}},\;
\end{equation}
where the parameter $\sigma$ characterizes the fractional helicity of $\ff$, and
\begin{equation}
\fh({\kk})^{\rm(nohel)}=
\left(\kk\times\eee\right)/\sqrt{\kk^2-(\kk\cdot\eee)^2}
\label{nohel_forcing}
\end{equation}
is a nonhelical forcing function.
We use only those $\eee$ that are not aligned with $\kk$.
Note that $|\fh|^2=1$.
We consider both $\sigma=0$ and $\sigma=1$,
corresponding to the nonhelical and maximally helical cases.
The forcing is only enabled during the time interval $0\leq t\leq t_*$.
In this sense, this forcing procedure can be considered as part of the initial condition.

In this section and henceforth, we use $t$ to refer to conformal time, as opposed
to coordinate time in Sec.~\ref{secModel}.
All other quantities are comoving
quantities, scaled by exploiting the conformal symmetry of Maxwell's equations; see
\cite{Brandenburg:1996fc} for details.
We solve Eqs.~(\ref{dlnrhodt})--(\ref{dAdt}) using the {\sc Pencil Code}, a public MHD code\footnote{\url{https://github.com/pencil-code}}, which is well suited for studying and simulating turbulence.
The simulations are performed in a periodic domain of size $L$.
Except for the homogeneous imposed magnetic field at wave number $k=0$,
the smallest nonvanishing wave number in the domain is $k_1\equiv2\pi/L$.
Spatial derivatives are computed using sixth order accurate finite
differences and a third order accurate time stepping scheme is used.
The magnetic vector potential is advanced in time to preserve
solenoidality (the divergence-free condition) of the magnetic field.
We use a numerical resolution of $1152^3$ meshpoints for all simulations
presented in this paper.

\subsection{Peculiarities connected with imposed fields}

In a periodic domain, the case of an imposed magnetic field is in many
ways pathological,
since it will always be present and can never decay.
It can be amplified linearly in time by a flow -- even in two dimensions
where no dynamo effect is possible \cite{Mos70}.
In addition, magnetic helicity associated with the induced magnetic field
based on the deviations of the magnetic
field from the imposed field is not conserved \cite{Ber97}.
This is because it interacts with the imposed field, which, owing to its
constancy in space, 
cannot have magnetic helicity.
On the other hand, if we replace the imposed field by a large-scale
field with zero net flux, the magnetic helicity becomes well defined.
The total field can then decay to zero, and the magnetic helicity is now
a perfectly defined quantity that obeys the usual conservation law.
We can therefore ask how the presence of a large-scale magnetic field
affects the evolution of magnetic helicity of a field of much smaller
length scale.

To better understand the aforementioned peculiarities,
we note that in the presence of an imposed magnetic field,
a generalized quantity can be defined that is still conserved \cite{MG82},
but that quantity is not gauge invariant and hence not uniquely defined
\cite{Brandenburg:2003pe}.
Let us discuss this here in more detail.
In the presence of an imposed field, $\mathbf{B}_0=\mbox{const}$, one splits
the magnetic field into a mean and a fluctuating component,
$\mathbf{B}=\mathbf{B}_0+\mathbf{b}$.
The mean of $\mathbf{b}$ is vanishing.
Using $\mathbf{b}=\mathbf{\nabla}\times\mathbf{a}$,
the time derivative of the volume-averaged quantity
$\langle\mathbf{a}\cdot\mathbf{b}\rangle$,
is found to have a term $-2\alpha\mathbf{B}_0^2$,
in addition to the Spitzer term $-2\eta\langle\mathbf{j}\cdot\mathbf{b}\rangle$;
see Appendix~A for the derivation.
Here, $\alpha$ refers to the $\alpha$ effect and it models the component of the electromotive
force, $\bm{\mathcal{E}}=\langle \mathbf{u}\times\mathbf{b}\rangle$, parallel to the
mean magnetic field.
The $\alpha$ effect is responsible for the fact that the mean magnetic helicity density
$\HHM=\langle\mathbf{a}\cdot\mathbf{b}\rangle$ is no longer conserved \cite{Ber97}.

The presence of an imposed magnetic field was found to influence the sign
of the magnetic helicity and the inverse cascade \cite{Haugen:2004zh}.
For weak (or zero) imposed fields, magnetic helicity
and energy cascade strongly from the forcing scale to large length scales,
and the magnetic helicity has an opposite sign to the kinetic helicity.
For stronger fields, the inverse cascade of magnetic helicity to larger scales is suppressed,
and the sign of the magnetic helicity flips over.
The threshold strength of the imposed magnetic
field depends inversely on the square root of the magnetic Reynolds number.
This is understood to be a consequence of the $\alpha$ effect.

It was also found that, in the
presence of an imposed magnetic field, the induced magnetic field can undergo a
certain enhancement around the forcing wave number.
Furthermore, small-scale dynamo action
helps to lower the energy density in the inertial region in $k$-space \citep{Haugen:2004zh}.
In addition, during the initial time interval $0\leq t\leq t_*$,
we drive turbulence either through the $\FFF_0$ or $\EEE_0$ terms.

\subsection{Initial conditions}

We consider the following types of initial conditions.
First, we consider a homogeneous (imposed) magnetic field \cite{Haugen:2004zh},
where the corresponding correlation length is infinite.
Second, we consider the case with no imposed field, but with a zero flux
initial scale-invariant magnetic field, so the correlation
length is finite,
but much longer than the scale of  turbulence.\footnote{
The finite value of the correlation length is determined by the cutoff scale 
imposed to the scale-invariant spectrum at the low wave numbers region 
\cite{Kahniashvili:2016bkp}} 
We construct such a field in Fourier space as
\begin{equation}
\tilde{B}_i({\bf k})=B_{\rm ini}\left(\delta_{ij}-{\hat k}_i {\hat k}_j-i\sigma\epsilon_{ijl} {\hat k}_l\right)
g_j({\bf k})\, |{\bf k}|^{-3/2},
\end{equation}
where ${\bf g}({\bf k})$ is the Fourier transform of a Gaussian distributed
random vector field that is $\delta$-correlated in all three dimensions.
The degree of helicity is controlled by the parameter $\sigma$,
which is $\pm1$ for maximally helical fields with positive
or negative helicity, and zero in the nonhelical case.
The magnetic field in real space is given by
$\BB(\xx)=\int\tilde{\BB}(\kk)e^{i\kk\cdot\xx}d^3k/(2\pi)^3$.
In all cases, we have initially $\rho=\mbox{const}$.

The forcing applied during $0\leq t\leq t_*$ consists of monochromatic
forcing (see, for example, Ref.~\cite{Kahniashvili:2010gp}) at a wave number $k=k_\star$.
This forcing wave number corresponds to a fraction of
the Hubble scale $H_\star$ after inflation.
One can think of this as the epoch of reheating.

In either case, we consider the relativistic fluid to have an initial turbulent velocity field $\mathbf{u}(\mathbf{x})$.
Physically, turbulence can be induced at reheating by energy injection from the inflaton into the
standard model particles and fields, or from bubble collisions during some (yet unknown)
phase transition -- the spectral energy density
$\EM(k)$ has a $k^4$ subinertial range at large scales due to causality requirements (see Refs.~\cite{Durrer:2003ja,Brandenburg:2018ptt}),
while in the inertial range, it tends to have a Kolmogorov spectrum proportional to $k^{-5/3}$.

\begin{table*}[t!]\caption{
Summary of the parameters of the simulations discussed in this paper.
Some characteristic parameters, including the final values of $p$ and $q$
as defined below and the direction of evolution in the $pq$ diagram are
also indicated.
}\vspace{12pt}\centerline{\begin{tabular}{ccccccccclccl}
panel & initial field & $B_0$ & $B_{\rm ini}$ & $v_{\rm A}^{\max}$ & ${\cal F}_0$ & ${\cal E}_0$ &
$\sigma$ & $k_*$ & $~\tau_{\rm A}$ & $p$ & $q$ & $\quad$ evolution along \\
\hline
(i) & homogeneous & 0.1 & 0         &0.08& 0.02 & 0   & 0 & 60 & 0.19  & 10/7 & 2/7 &$\quad\beta=4$ to $(p,q)\to(10/7,\,2/7)$ \\%1152a
(ii)& homogeneous & 0.1 & 0         &0.08& 0.02 & 0   & 1 & 60 & 0.19  & 10/7 & 2/7 &$\quad\beta=4$ to $(p,q)\to(10/7,\,2/7)$ \\%H1152a
(a) & homogeneous & 0.03& 0         &0.46& 0 & 0.0005 & 1 &180 & 0.21  &  2   &  0  &$\quad\beta=4$ to $(p,q)\to(2/3,\,2/3)$  \\%I1152kf180f1
(b) & homogeneous & 0.10& 0         &0.41& 0 & 0.0005 & 1 &180 & 0.06  &  2   &  0  &$\quad p=2(1-q)$ to $(p,q)\to(2,0)$      \\%I1152kf180a2
(c) & homogeneous & 0.16& 0         &0.31& 0 & 0.0005 & 1 &180 & 0.04  &  4   &  0  &$\quad\beta=1$--$2$ to $(p,q)\to(1,\,0.5)$  \\%I1152kf180k1
(d) & homogeneous & 0.20& 0         &0.25& 0 & 0.0005 & 1 &180 & 0.03  &  4   &  0  &$\quad\beta=3$--$4$ to $(p,q)\to(0.1,\,0.8)$ \\%I1152kf180j1
(e) & homogeneous & 1.00& 0         &0.21& 0 & 0.0005 & 1 &180 & 0.006\;& 4   &  0  &$\quad\beta=3$--$4$ to $(p,q)\to(0,0)$  \\%I1152kf180b3
(A) & $1/k$ spectrum & 0   & $10^{-3}$ &0.47& 0 & 0.0005 & 1 &180 & 6.4~  & 0.6  & 0.6 &$\quad\beta=0$ to $(p,q)\to(0.6,\,0.6)$ \\%J1152kf180c
(B)&$1/k$ spectrum &0& $3\times10^{-2}$&0.37& 0 & 0.0005 & 1 &180 & 0.22  & 0.2  & 0.2 &$\quad\beta=0$ to $(p,q)\to(0.2,\,0.2)$ \\%J1152kf180a2
\label{Summary}\end{tabular}}\end{table*}

We recall that, in the absence of a large-scale magnetic field, a small-scale helical
magnetic field undergoes inverse cascading such that the magnetic energy
at small wave numbers increases with time \cite{Christensson:2000sp,Banerjee:2004df}.
The characteristic length scale of the turbulence, $\xiM$,
increases with time like $t^{2/3}$, and the magnetic energy $\EEM$
decreases like $t^{-2/3}$, which is slower than in the nonhelical
case where $\EEM\propto t^{-1}$ and $\xiM\propto t^{1/2}$.

One often considers the magnetic field evolution in a diagram
of $\EEM$ versus $\xiM$.
The nonobservation of GeV cascade photons from the interaction of
TeV photons from blazars with the extragalactic background light,
as mentioned above, has
often been argued to imply the presence of a lower limit on the
product $\EEM\xiM$ of about $(10^{-15}\G)^2\Mpc$.
In a diagram of $\EEM$ versus $\xiM$, the line corresponding to this lower limit
has a slope of $-1$, which is also the slope of the line representing
the magnetic field decay in the fully helical case, because
$\EEM\propto t^{-2/3}\propto\xiM^{-1}$.
For a nonhelical field, on the other hand, we have
$\EEM\propto t^{-1}\propto\xiM^{-2}$.
For this reason, a nonhelical field will eventually drop below the line
demarcating the lower observational limit \cite{Brandenburg:2017neh}.
We now study how these decay properties are affected by the presence of either
an imposed or an initial large-scale magnetic field.

\begin{figure*}[t!]
\begin{center}
\includegraphics[width=\columnwidth]{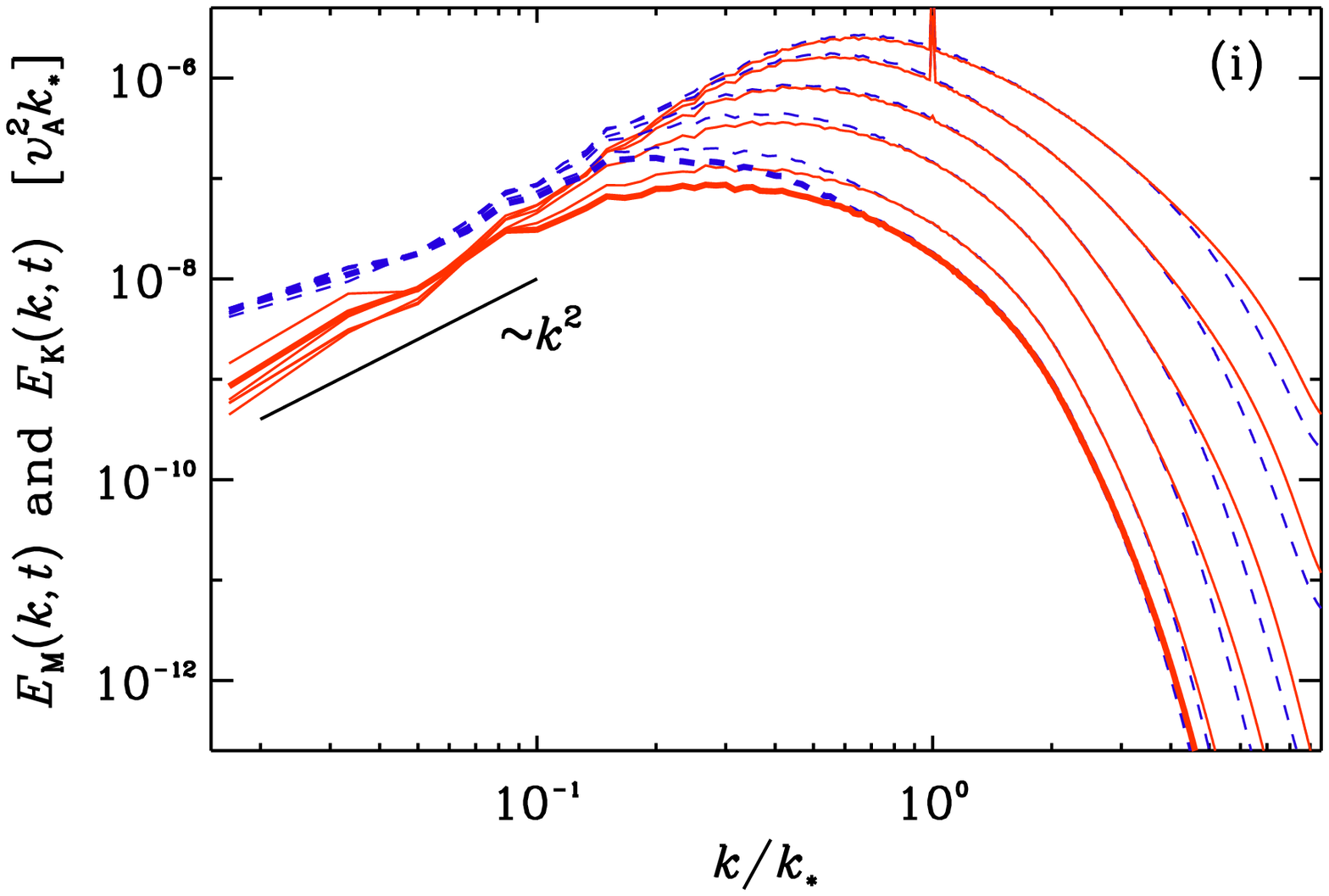}
\includegraphics[width=\columnwidth]{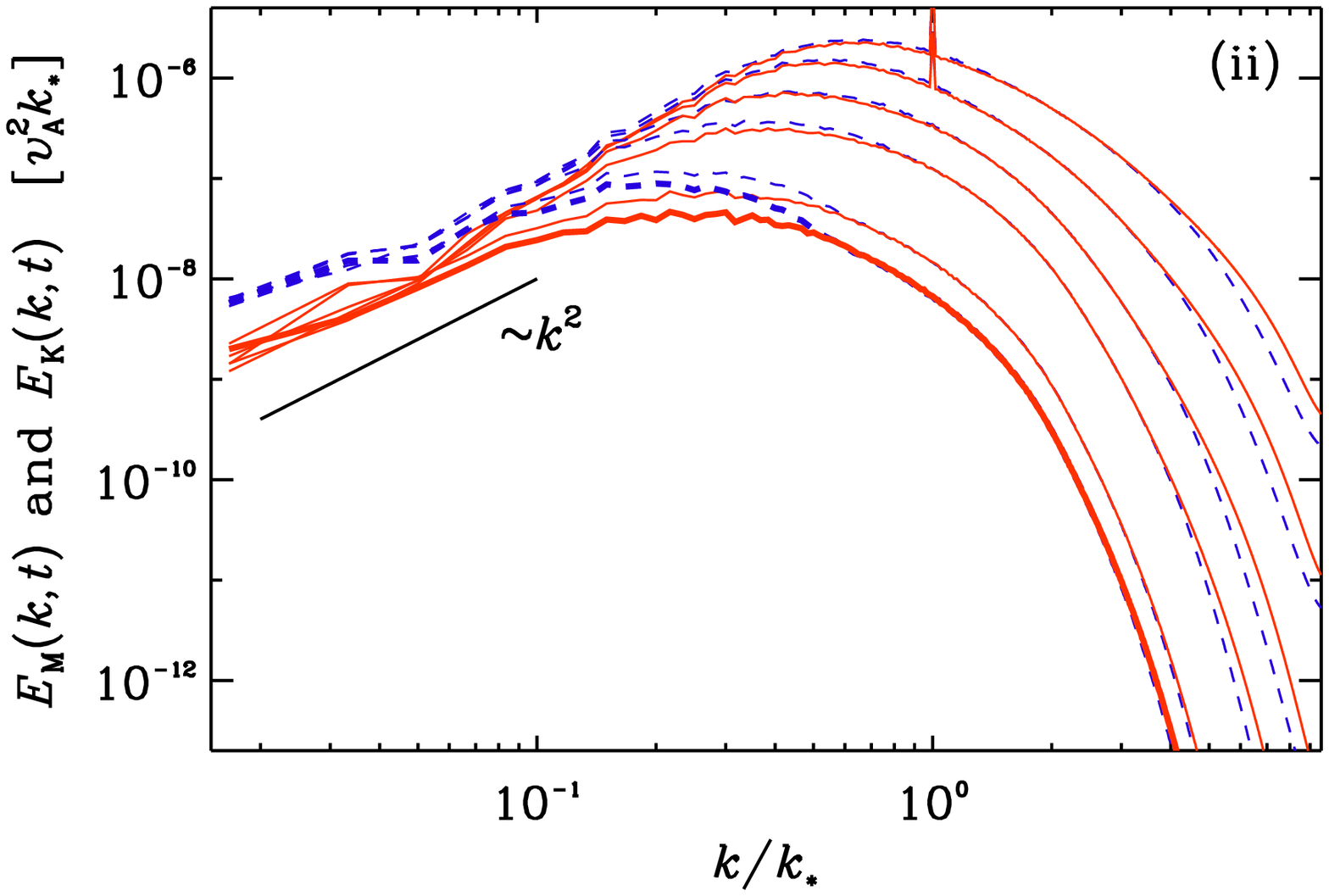}
\end{center}
\caption[]{The evolution of the magnetic (red) and kinetic (blue) energy
spectra for (i) nonhelical and (ii) helical turbulence.
The thick lines are the configurations at the latest times.
Panels (i) and (ii) correspond to runs~(i) and (ii) in \Tab{Summary}.
}\label{pkt1152_1152a}
\end{figure*}

\subsection{Parameters and analysis tools}

By default, we measure lengths in units of $k_1^{-1}=L/2\pi$
and wave numbers in units of $k_1$.
Since $c=1$, time is measured in units of the light travel time,
$(ck_1)^{-1}$, and viscosity or magnetic diffusivity are measured
in units of $c/k_1$.
Furthermore, since $\rho=1$ initially, the magnetic field is
measured in units of $c/\sqrt{\rho}$.
Our main control parameters are $k_*$, the amplitudes
of the imposed or initial fields, $B_0$ and $B_{\rm ini}$, respectively,
the amplitudes of the forcing functions ${\cal E}_0$ and ${\cal F}_0$,
and the values of $\nu$ and $\eta$.
For $k_*$, we consider the values 60 and 180,
$B_0$ and $B_{\rm ini}$ are varied between 0 and 1,
while ${\cal E}_0$ and ${\cal F}_0$ are varied 0 and 0.02,
such that the energy density of the turbulence does not exceed the radiation
energy density by more than 10\% after the duration of turbulent driving,
which we have chosen to be $t_*=5$ in the normalized units defined below.
In all cases, we use a resolution of $1152^3$ meshpoints
and we found that $\nu=\eta=10^{-5}$ is sufficiently small to
dissipate the energy of the turbulence at the smallest length scale.
A summary of parameters of all runs is given in \Tab{Summary}.

\begin{figure*}[t!]
\begin{center}
\includegraphics[width=.99\columnwidth]{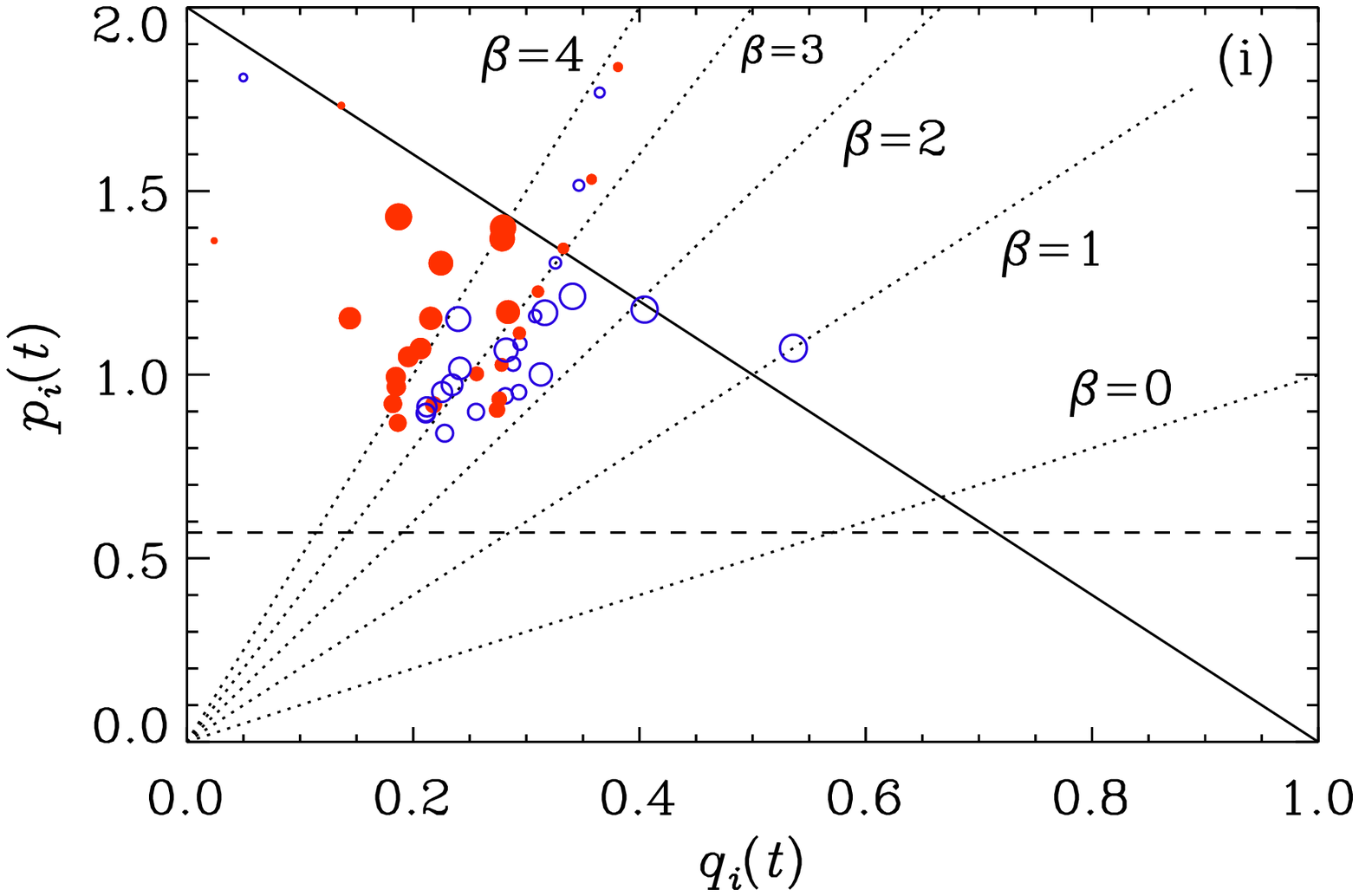}
\includegraphics[width=.99\columnwidth]{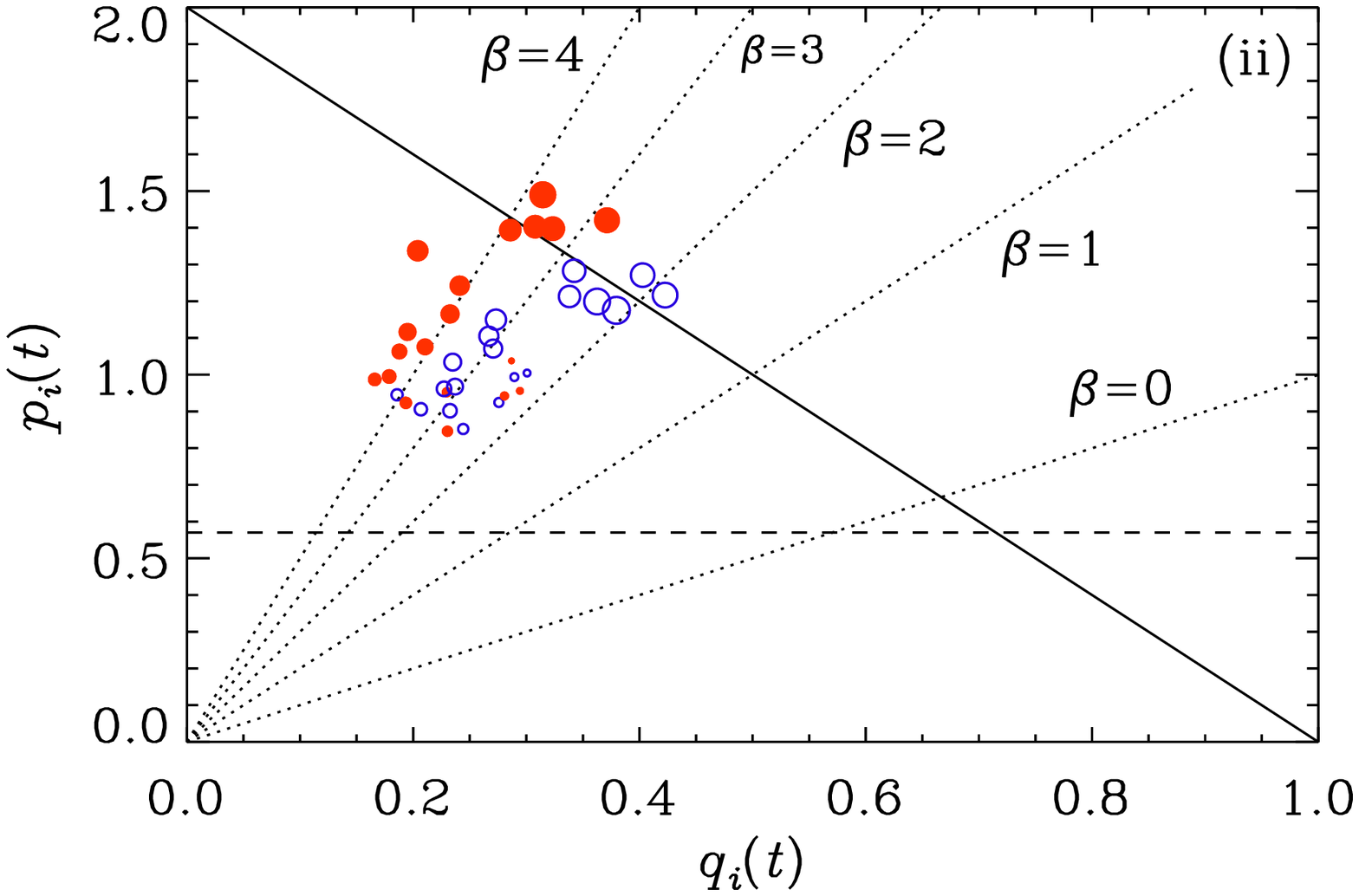}
\end{center}
\caption[]{$pq$ diagrams for the cases of nonhelical (i) and helical (ii) turbulence.
The dots denote the instantaneous values of $(p,q)$ for the magnetic (red) and kinetic (blue) fields.
Bigger circles denote later times.
Panels (i) and (ii) correspond to runs~(i) and (ii) in \Tab{Summary}.
The solid lines correspond to the self-similarity line, $p=2(1-q)$, the
dotted lines denote $\beta=\mbox{const}$, and the dashed lines denote
$p=\mbox{const}\approx0.58$, whose relevance is explained in the text.
}\label{pq_1152a}
\end{figure*}

It is sometimes convenient to express time in units of the Alfv\'en time,
$\tau_{\rm A}=(v_{\rm A} k_*)^{-1}$, where
$v_{\rm A}^2=B_0^2/(\fourthird\rho)$ for
cases with an imposed magnetic field and $v_{\rm A}^2=B_{\rm ini}^2/(\fourthird\rho)$
for cases with a zero flux large-scale magnetic field.
To specify the strength of the fluctuating magnetic field in cases with
$B_0\neq0$, we also specify the quantity
$v_{\rm A,f}^{\max}=|\BB-\BB_0|/(\fourthird\rho)^{1/2}$.
The kinetic and magnetic energy densities are defined as
$\EEK=\bra{\rho\uu^2}/2$ and $\EEM=\bra{\BB^2}/2$, respectively,
and the kinetic and magnetic energy spectra,
$\EK(k,t)$ and $\EM(k,t)$, are normalized such that
\begin{equation}
\int \EK(k,t)\,dk=\EEK\quad\mbox{and}\quad
\int \EM(k,t)\,dk=\EEM,
\end{equation}
respectively.
We define the magnetic correlation length $\xiM$ as
\begin{equation}
\xiM(t)=\left.\int k^{-1}\EM(k,t)\,dk\right/\!\!\int \EM(k,t)\,dk.
\end{equation}
Finally, we define the instantaneous exponents describing
the growth of $\xiM(t)$ and the decay of $\EEM(t)$ as
\begin{equation}
q_i(t)=d\ln\xi_i/d\ln t,\quad
p_i(t)= \rut{-} d\ln\EEi/d\ln t.
\end{equation}
Those play important roles in describing the nature of the turbulence
in different cases \cite{Brandenburg:2016odr}.

The various solutions are characterized by certain lines in the
$pq$ diagram.
It was found that the point $(p,q)$ ultimately settles somewhere on what
was called the self-similarity line \cite{Brandenburg:2016odr}, where
\begin{equation}
p=2(1-q).
\label{selfsim}
\end{equation}
Moreover, this evolution occurs along a line with
\begin{equation}
\beta=p/q-1=\mbox{const},
\label{betaconst}
\end{equation}
where the value of $\beta$ is determined by the nature of certain
relevant conservation laws.
Eliminating $p$ from Eqs.~(\ref{selfsim}) and (\ref{betaconst}), we find
$\beta=2/q-3$, where $q$ can be obtained from dimensional arguments
in terms of the dimensions of length $L$ and time $T$.
We recall that $q$ characterizes the scaling of the correlation length
with time as $\xiM\sim t^q$.
Magnetic helicity has dimensions $L^3 T^{-2}$, so $q=2/3$,
and therefore $\beta=0$.
The mean squared vector potential, which is arguably relevant to
magnetically dominated turbulence \cite{Brandenburg:2014mwa}, has
dimensions $L^4 T^{-2}$, so $q=1/2$, and therefore $\beta=1$.
The Saffman integral \cite{Saf67} has dimensions $L^5 T^{-2}$, so $q=2/5$,
and therefore $\beta=2$, while the Loitsiansky integral \cite{BP56}
has dimensions $L^7 T^{-2}$, so $q=2/7$, and therefore $\beta=4$.

Under certain conditions, the evolution may not be self-similar for
extended periods of time.
In fact, for finite resolution and finite domain size, a truly
self-similar behavior is generally difficult to obtain.
A prolonged evolution along the line $p=\mbox{const}\approx0.58$
was obtained \cite{Brandenburg:2017rnt} when there is a complex interplay
between kinetic and current helicities.
In the present work, we find examples of several of the aforementioned
relations.

\section{Results}
\label{secResults}

\subsection{Helical and nonhelical decay with imposed field}

We consider decaying turbulence produced during a short initial period through
forcing at small scales with $k_*=60$ together with an imposed magnetic field.
We find that in the subinertial range, the magnetic energy spectrum goes approximately as $k^2$,
while the kinetic energy spectrum is shallower.
In Figs.~\ref{pkt1152_1152a}(i) and \ref{pkt1152_1152a}(ii), we show the evolution
of the magnetic and kinetic energy spectra for the nonhelical and helical cases,
respectively.
We see that the winding up of the initially uniform field by turbulence
causes a Saffman spectrum for the magnetic energy of the form $\EM\sim k^2$,
which is shallower than the Batchelor $k^4$ spectrum.
There is no inverse cascade in the sense that, even at small $k$,
the magnetic energy always decays.
The decay is faster at larger $k$, which causes $\xiM$ to increase,
but this is not due to the usual inverse cascade.

\begin{figure*}[t!]
\begin{center}
\includegraphics[width=\textwidth]{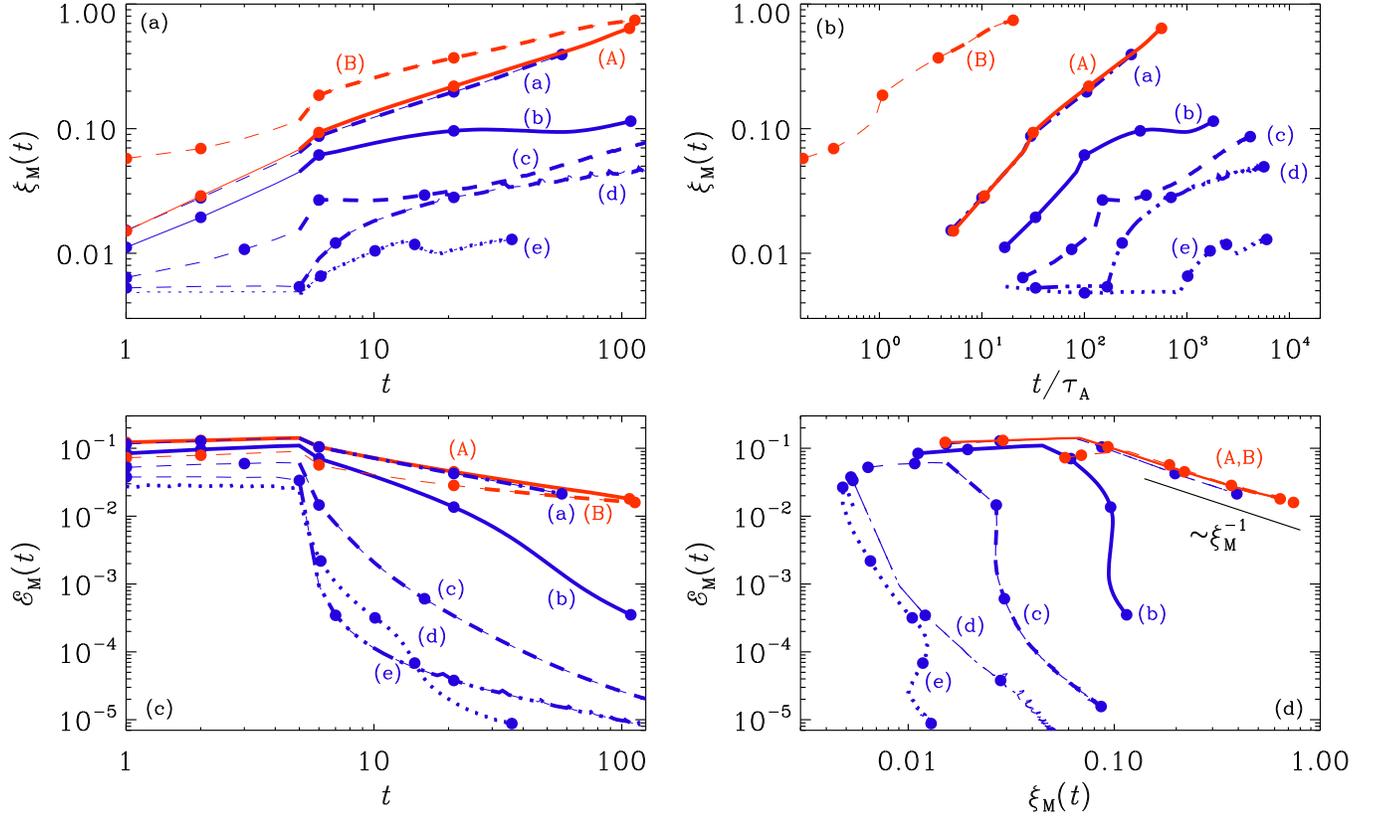}
\end{center}
\caption[]{
Dependences of $\xiM(t)$, $\EEM(t)$, $\EEM(t)/v_{\rm A}$,
and $\EEM(t)$ against $\xiM(t)$.
In the last panel, the slope $\EEM\propto\xiM^{-2/3}$
is shown for comparison.
The time interval $0\leq t\leq t_*\equiv5$ is marked with thin lines,
while later times are marked with thick lines.
Blue (red) lines denote cases with a perfectly homogeneous (statistically
homogeneous) magnetic field.
Solid (dashed) lines correspond to cases with a weak (strong) magnetic
field; compare labels (a)--(d) with the corresponding runs in \Tab{Summary}.
The filled symbols on each curve denote the five instances
for which the spectra below are shown.
}\label{pEkM2}
\end{figure*}

To quantify the decay further, we now show
in Figs.~\ref{pq_1152a}(i) and \ref{pq_1152a}(ii) the
evolution of the instantaneous scaling exponents
$p_i(t)$ versus $q_i(t)$ for $i={\rm M}$
and ${\rm K}$, where ${\mathcal E}_i$ is the energy density and $\xi_i$ is the
integral length scale for the magnetic and fluid fields.
We see that for both the helical
and the nonhelical cases, the evolution of the point $(p,q)$ tends to be close to the $\beta=4$ line, which implies the
conservation of the Loitsiansky integral \cite{BP56}.
This evolution is similar to that of nonhelical and nonmagnetic turbulence,
which is quite surprising: in the presence of a sufficiently strong
constant magnetic field, magnetic helicity seems to have no effect,
and the decay is very different from that in magnetically
dominated turbulence, where $\beta=1$--$2$ has been found
\cite{Brandenburg:2014mwa,Brandenburg:2016odr}.

\begin{figure*}[t!]
\begin{center}
\includegraphics[width=0.49\textwidth]{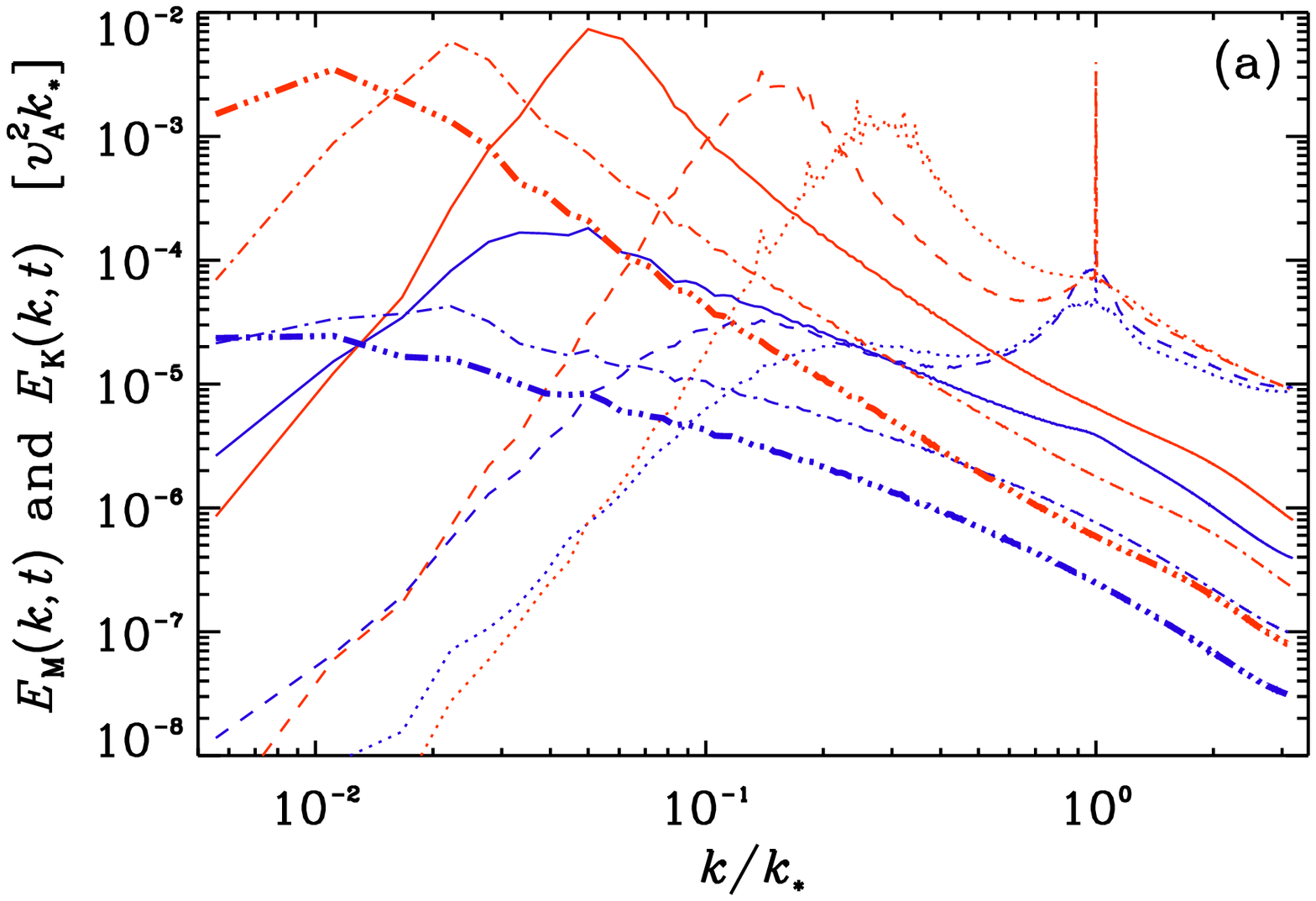}
\includegraphics[width=0.49\textwidth]{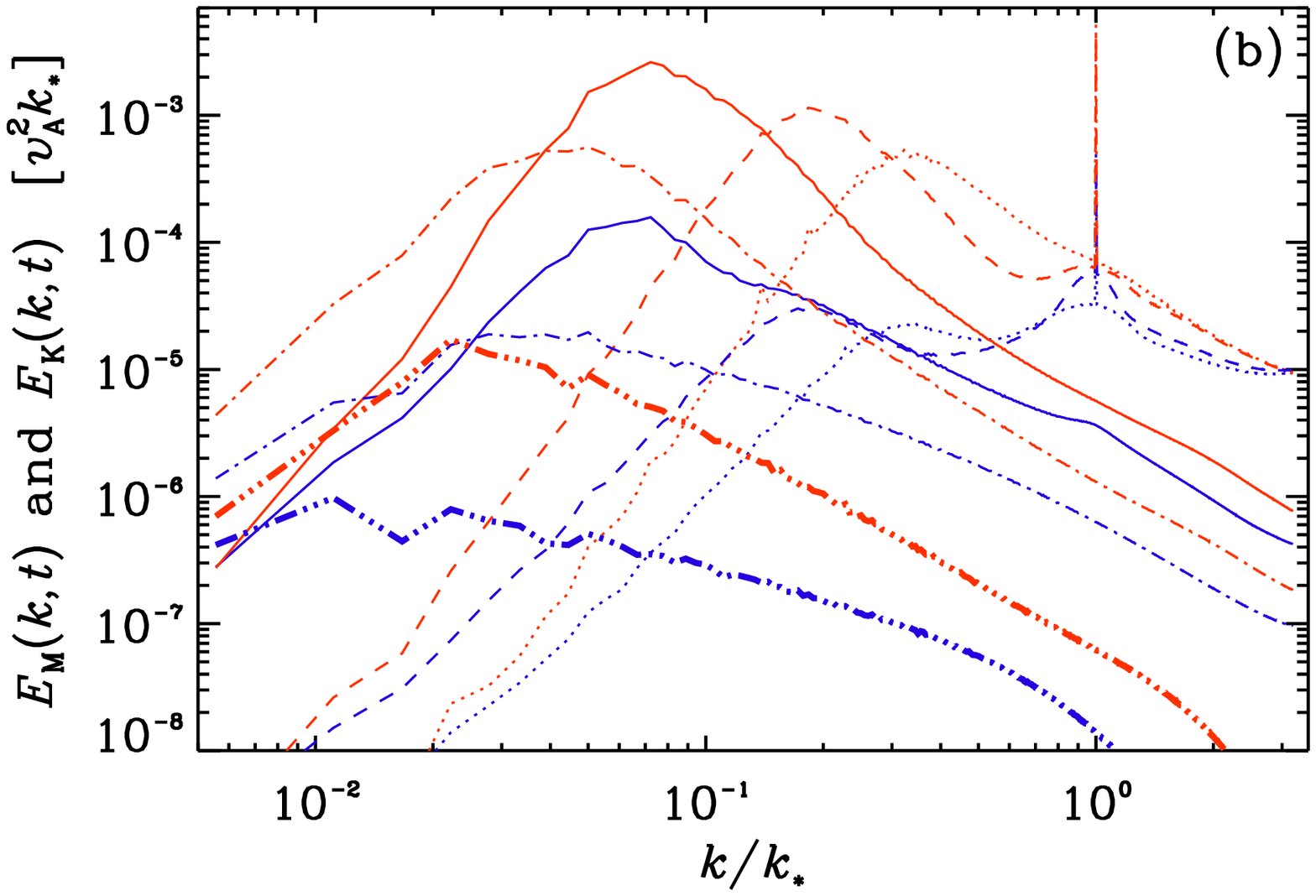}
\includegraphics[width=0.49\textwidth]{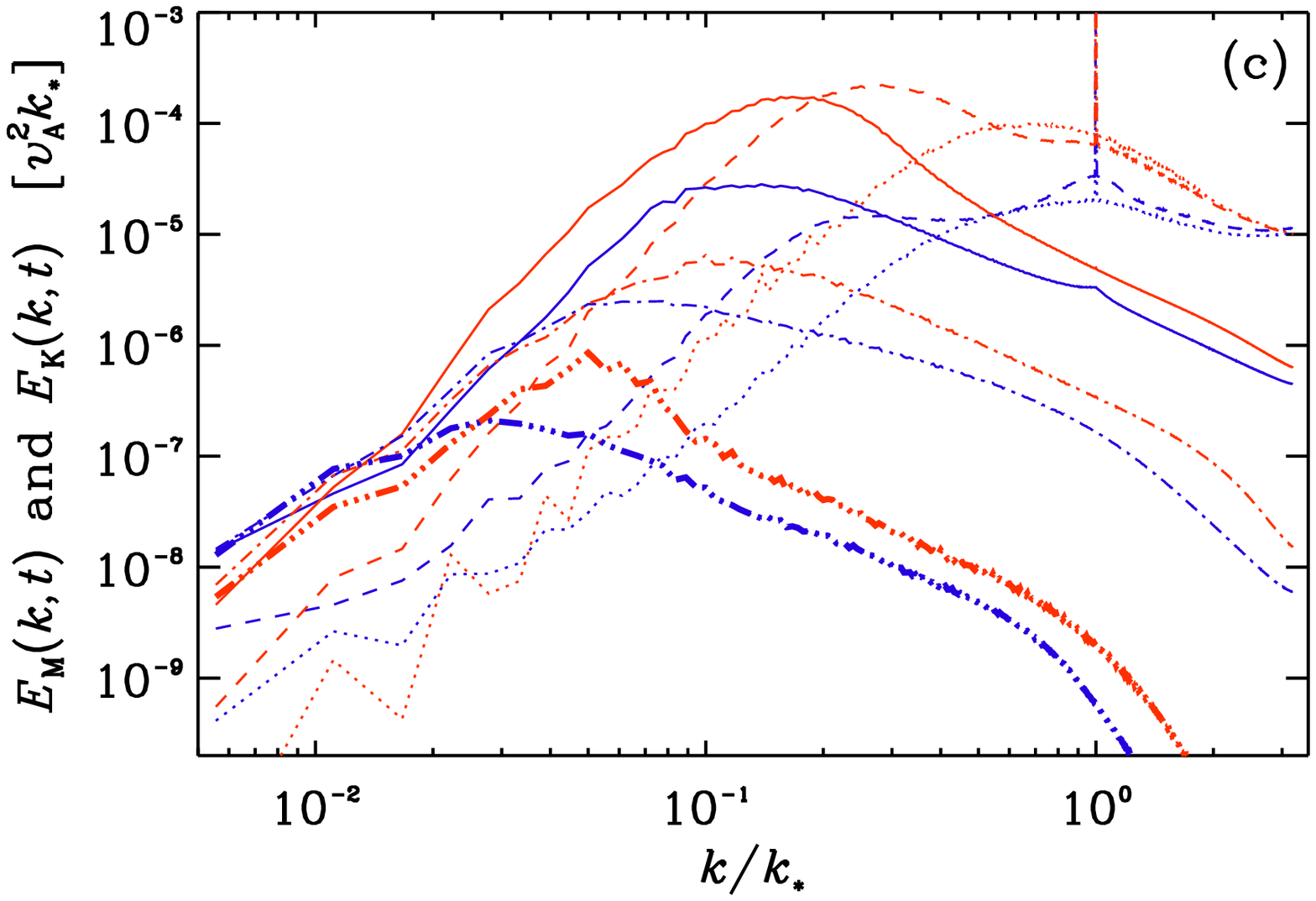}
\includegraphics[width=0.49\textwidth]{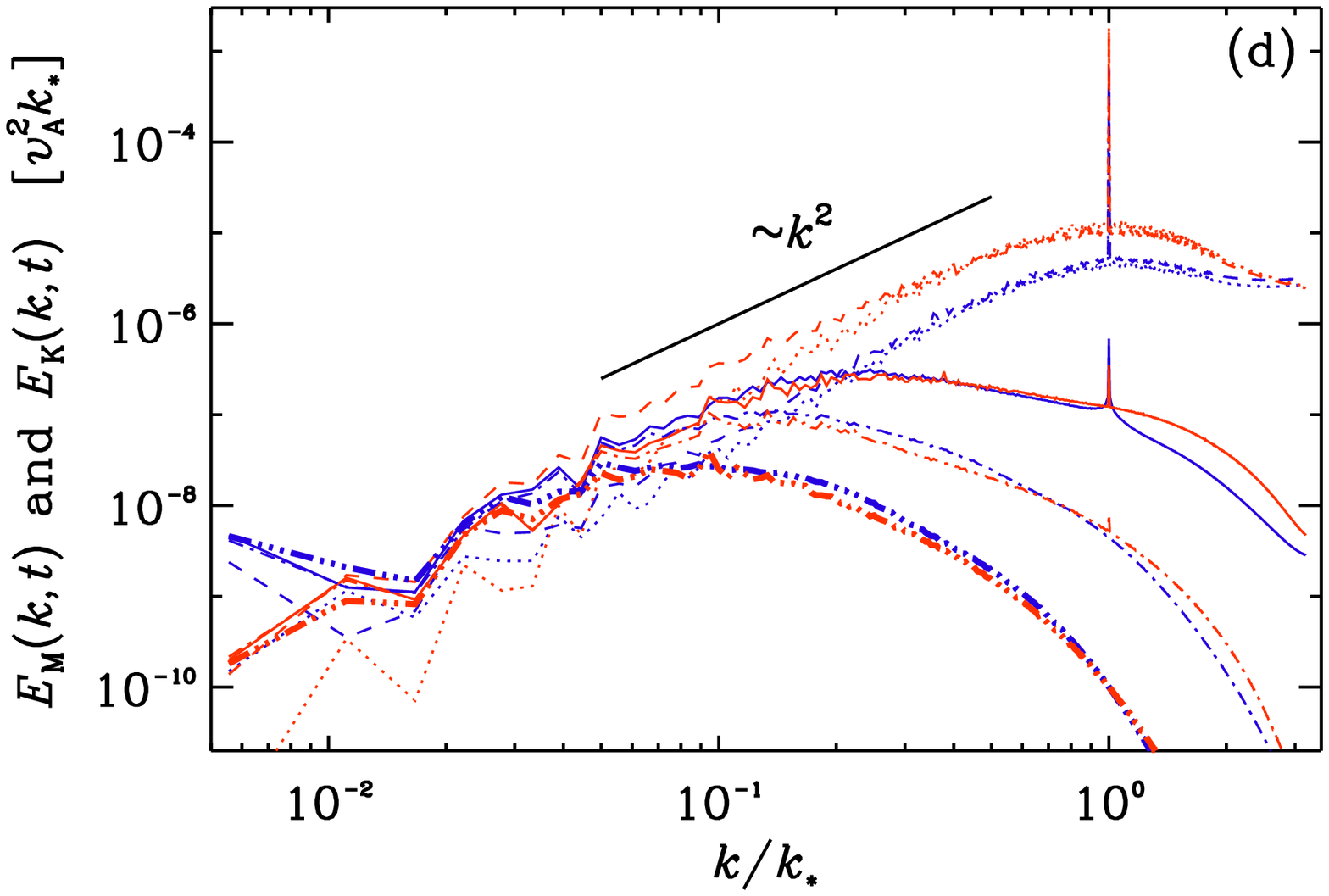}
\end{center}
\caption[]{
Magnetic (red) and kinetic (blue) energy spectra for
$k_*/k_1=180$ with an imposed field, $B_0=0.03$, $0.1$, $0.16$,
and $0.2$ in panels~(a)--(d), respectively.
These panels correspond to runs~(a)--(d) in \Tab{Summary}.
Dotted, dashed, solid, dash-dotted, and dash-triple-dotted lines
indicate later times, denoted by filled symbols in \Fig{pEkM2}.
The last time is also shown as a fat line.
}\label{pkt1152}
\end{figure*}

\begin{figure*}[t!]
\begin{center}
\includegraphics[width=0.49\textwidth]{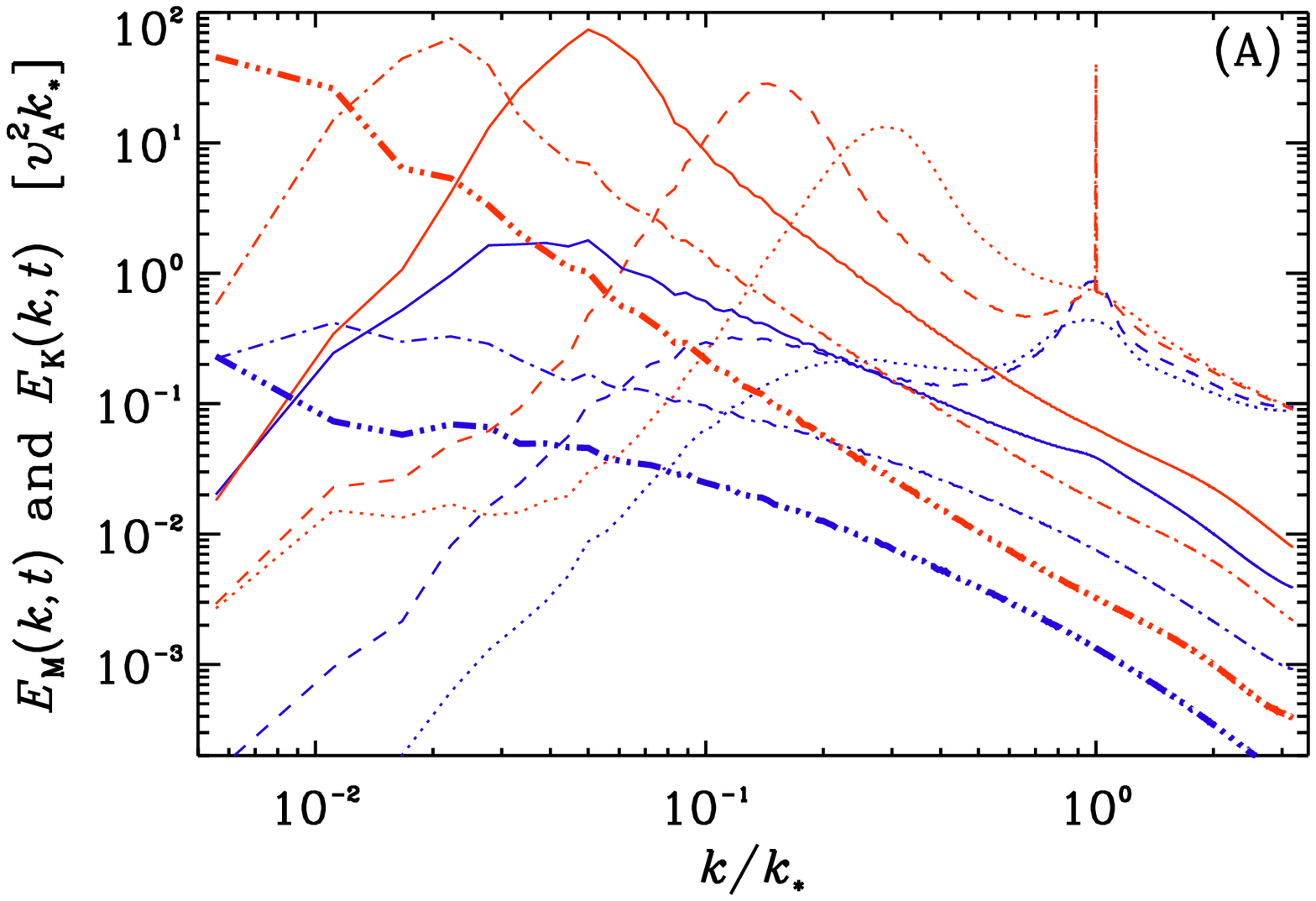}
\includegraphics[width=0.49\textwidth]{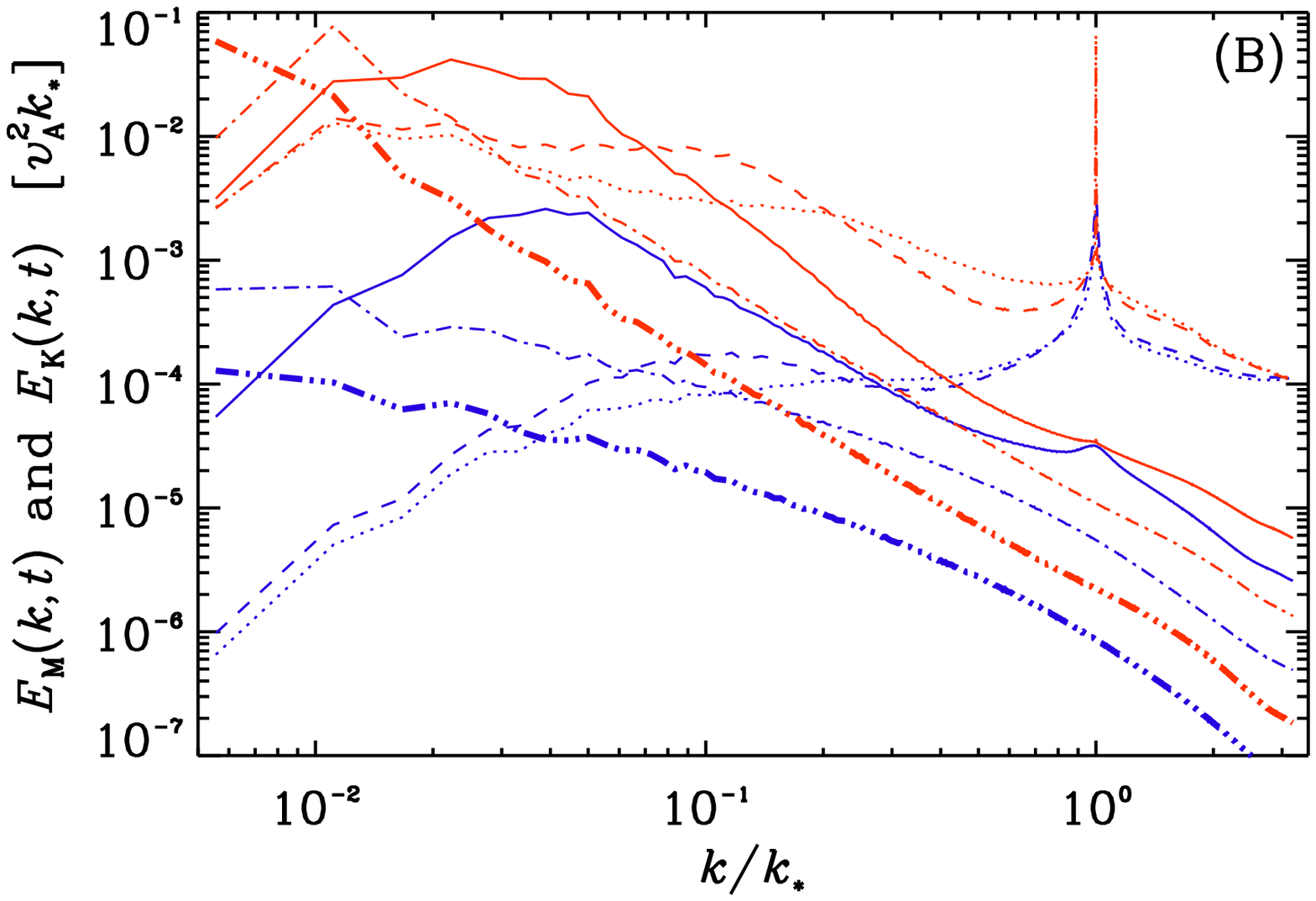}
\end{center}
\caption[]{
Similar to \Fig{pkt1152}, but for an initial large-scale field,
$B_{\rm ini}=10^{-3}$ and $3\times10^{-2}$ in panels~(A) and (B),
respectively.
}\label{pkt1152b}
\end{figure*}

\begin{figure*}[t!]
\begin{center}
\includegraphics[width=0.49\textwidth]{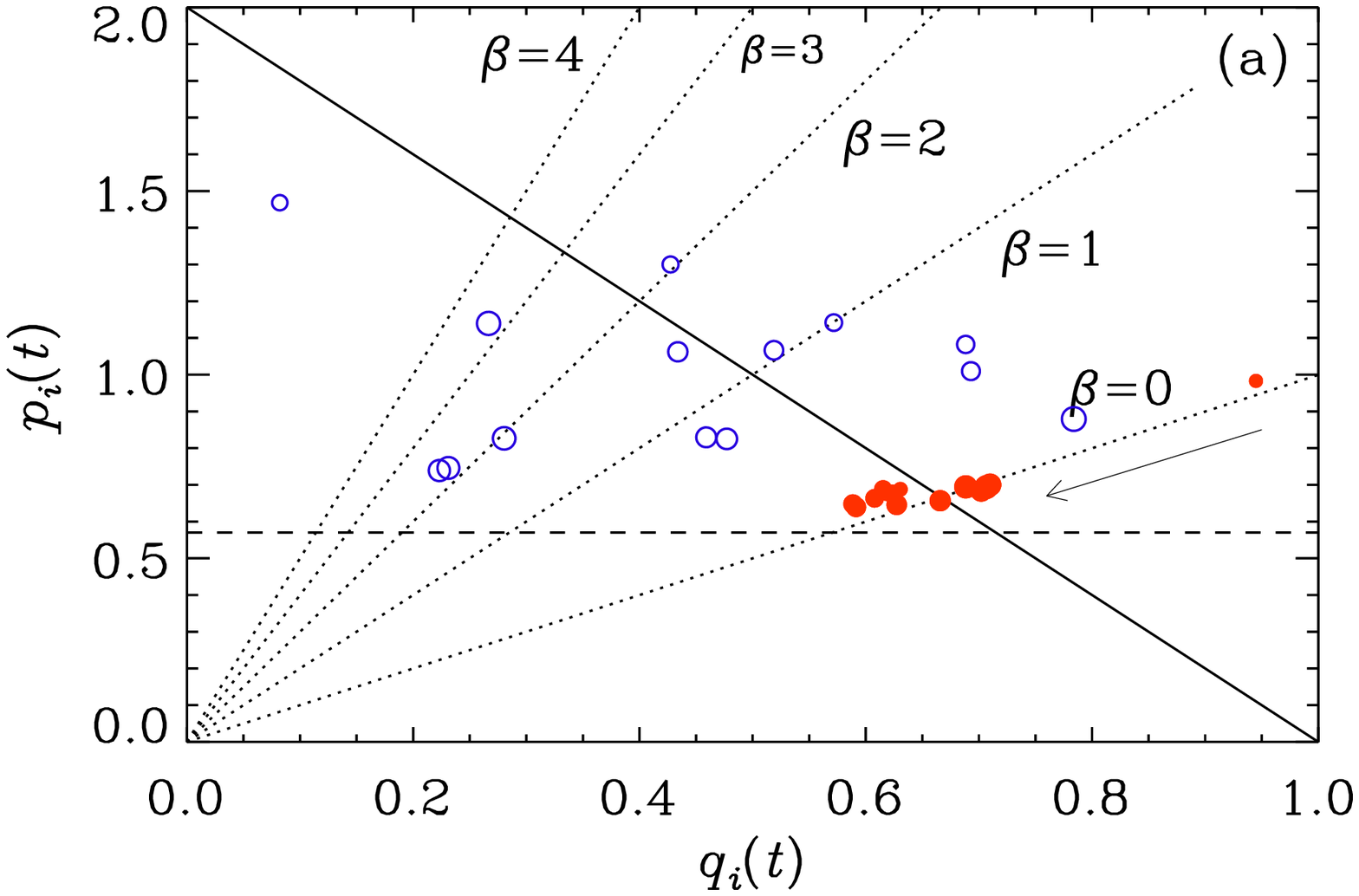}
\includegraphics[width=0.49\textwidth]{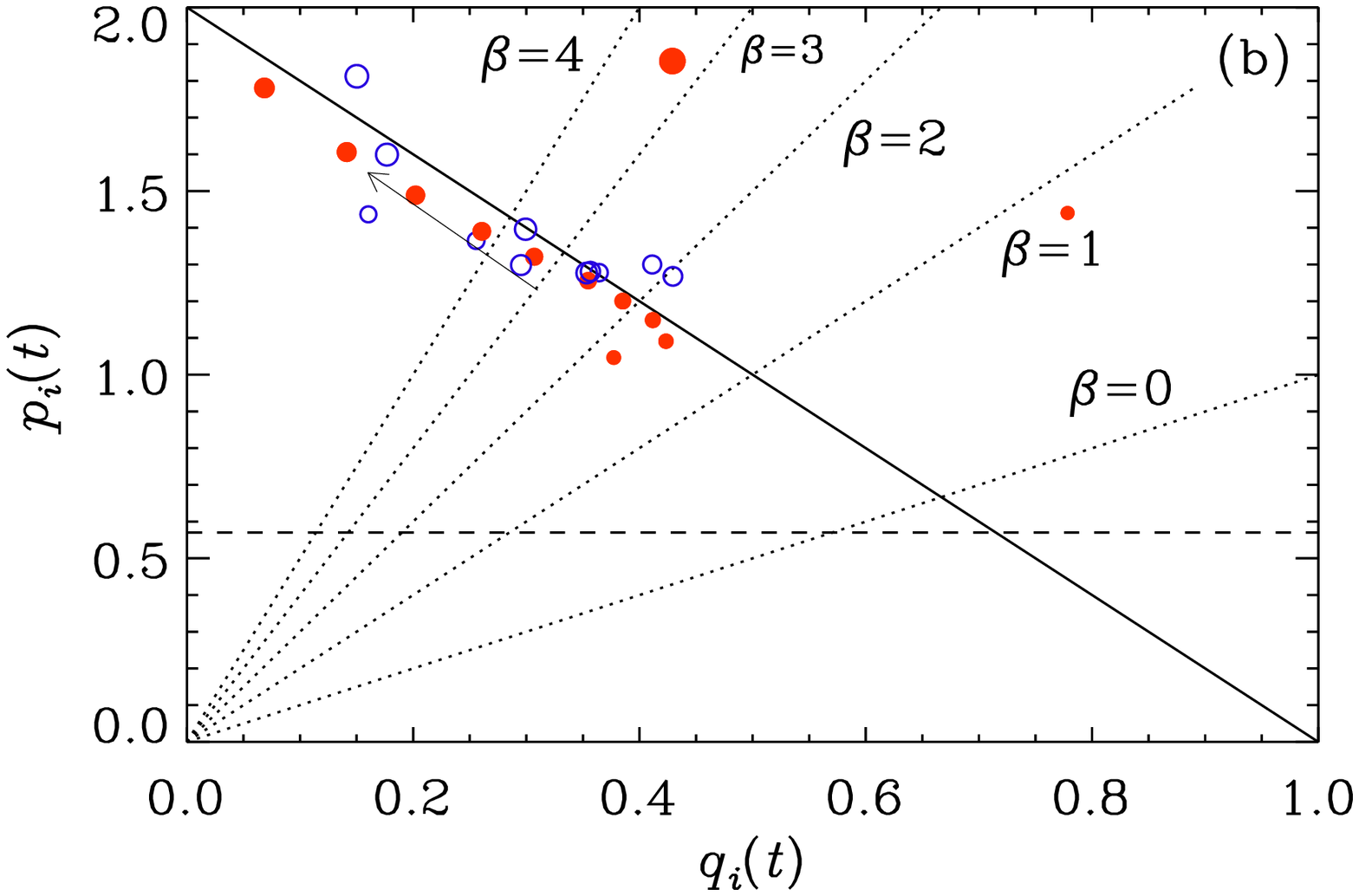}
\includegraphics[width=0.49\textwidth]{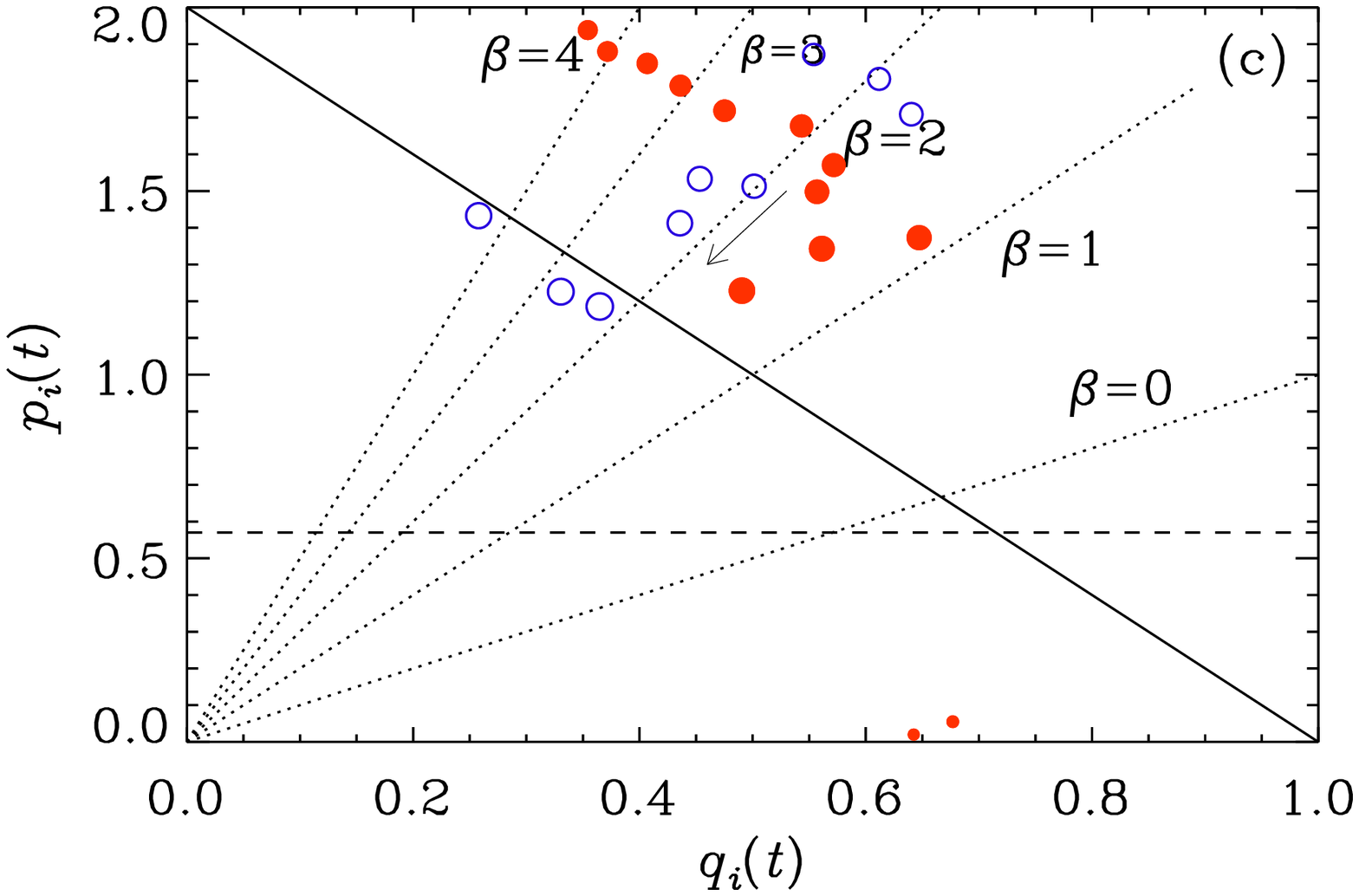}
\includegraphics[width=0.49\textwidth]{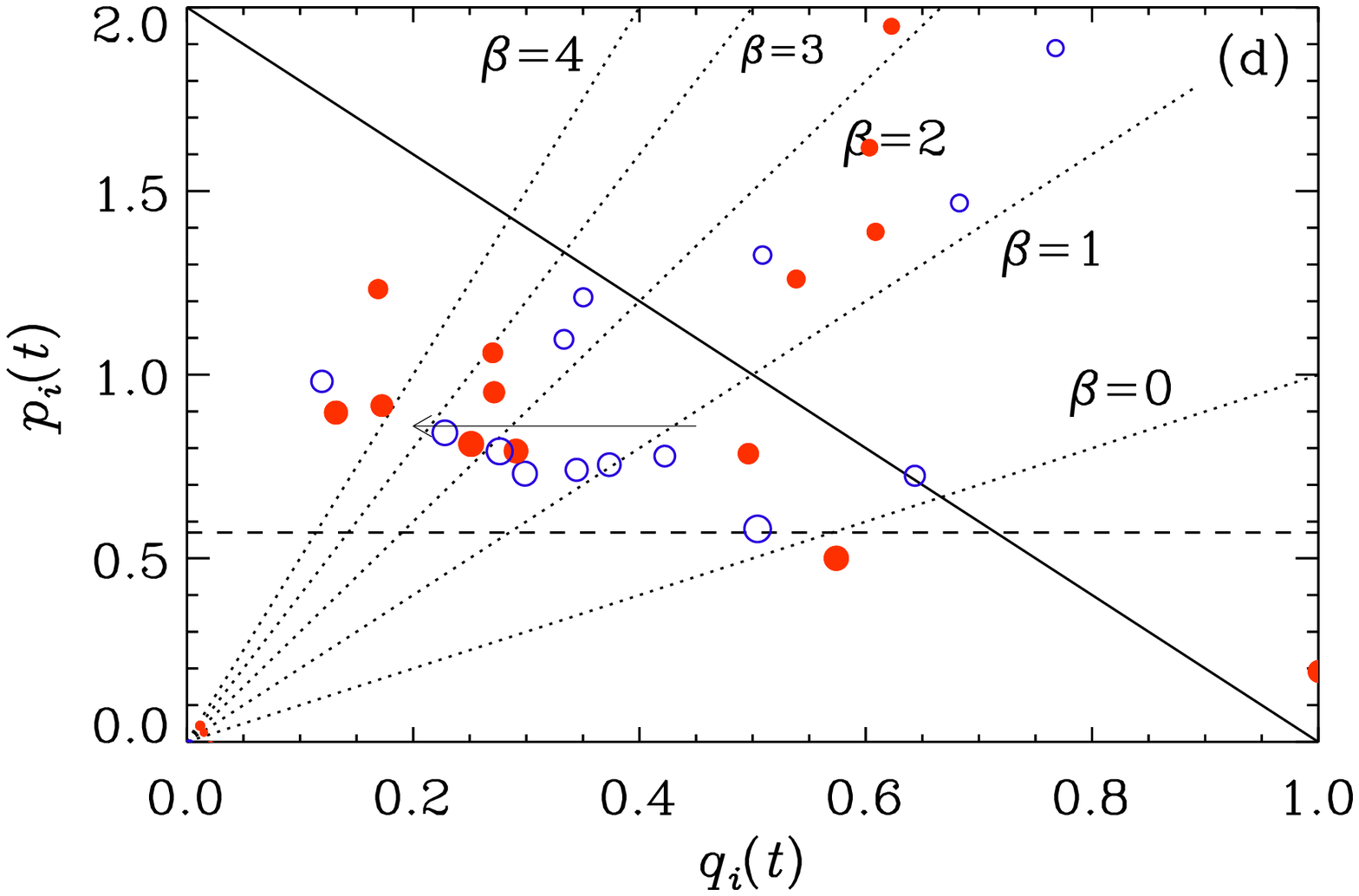}
\end{center}
\caption[]{
$pq$ diagrams for the magnetic field (red) and the velocity field (blue)
for $k_*/k_1=180$ with an imposed field, $B_0=0.03$, $0.1$, $0.16$,
and $0.2$ in panels~(a)--(d), respectively.
Again, these panels correspond to runs~(a)--(d) in \Tab{Summary}.
Later times are shown as larger symbols.
The arrows in each panel indicate the tentative direction of evolution.
}\label{pq}
\end{figure*}

\begin{figure*}[t!]
\begin{center}
\includegraphics[width=0.49\textwidth]{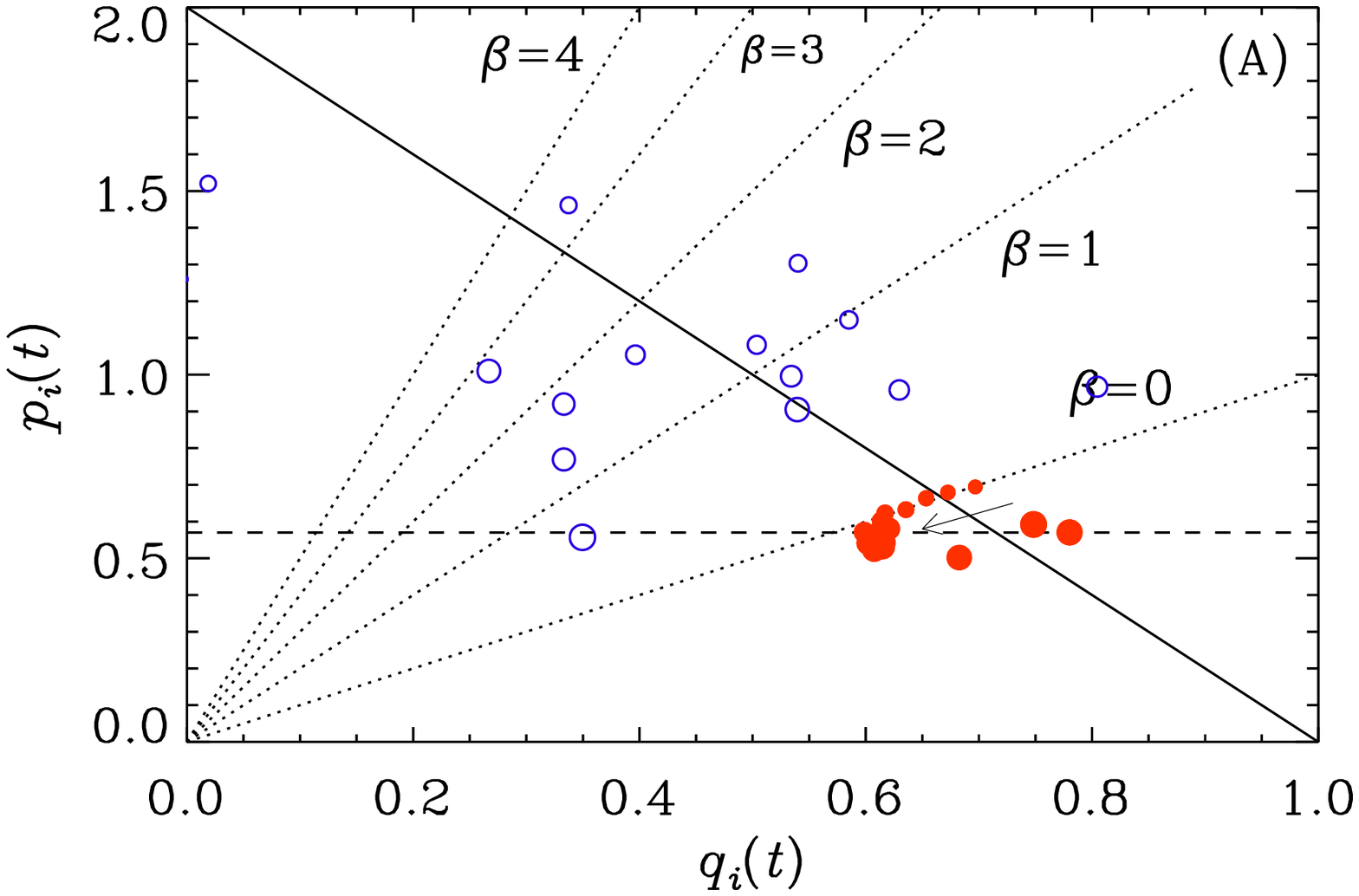}
\includegraphics[width=0.49\textwidth]{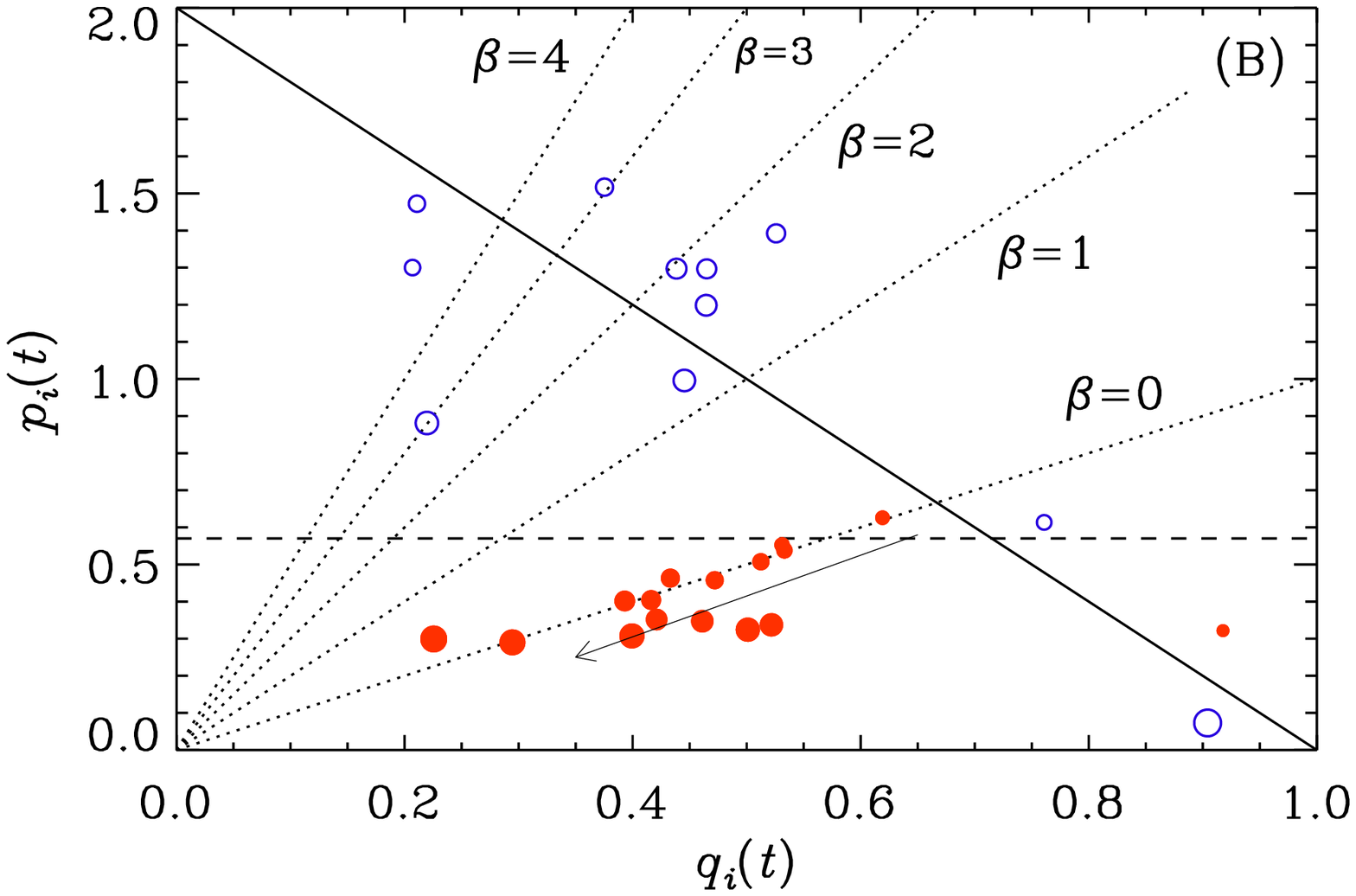}
\end{center}
\caption[]{
Similar to \Fig{pq}, but for an initial large-scale field,
$B_{\rm ini}=10^{-3}$ and $3\times10^{-2}$ in panels~(A) and (B),
}\label{pq2}
\end{figure*}

\subsection{Inverse cascade}

We know that, in the absence of a large-scale magnetic field, a
small-scale helical magnetic field decays more slowly than a nonhelical one, and
also its correlation length increases faster than for a nonhelical field.
It is therefore of interest to study how the magnetic decay is affected
by the presence of this large-scale magnetic field.
One may also ask whether some of the magnetic energy of this large-scale
field can be transferred to the smaller scale field.

In all cases, we produce a small-scale helical magnetic field by driving
the system with a turbulent small-scale electromotive force for a
short time interval $0\leq t\leq t_\ast=5$.
This driving is then turned off, leaving the system to decay freely,
except for the presence of the imposed magnetic field.
The runs are summarized in the lower part of \Tab{Summary}.

The time evolution of $\xiM(t)$ and $\EEM(t)$ is shown in
Figs.~\ref{pEkM2}(a) and \ref{pEkM2}(c).
In Fig.~\ref{pEkM2}(b) we plot the evolution $\xi(t/\tau_{\rm A})$
versus normalized time Fig.~\ref{pEkM2}(d).
Our results allow us to show $\EEM(t)$ against $\xiM(t)$ in a
parametric fashion; see Fig. \ref{pEkM2}.
Note that we have not included in $\EEM(t)$ the additional presence
of the imposed magnetic field, i.e., the magnetic energy is defined
solely based on the magnetic field with nonvanishing wave numbers.

In \Figs{pkt1152}{pkt1152b}, we present magnetic energy spectra for
cases with an imposed and an initial magnetic field, respectively.
In both cases, we see inverse cascading of the magnetic energy when the
imposed or initial magnetic fields are weak.
However, when the field is increased, the inverse cascade eventually
stalls; see especially \Fig{pkt1152}(c), where inverse cascading
has stopped after the peak of the spectrum traversed the $k$ axis by
about a factor of 10.

Interestingly, in the presence of an initial (nonimposed)
magnetic field, the evolution of $\EEM(t)$ versus $\xiM(t)$
follows the same line in \Fig{pEkM2}(d).
This line corresponds to $\EEM\propto\xiM^{-1}$ and its height
in that diagram characterizes the strength of magnetic helicity
\cite{Brandenburg:2017neh}.
Although the magnetic field was initially of small scale only,
at the end of the evolution, it has reached the scale of the system.
This is true for both weak and strong initial (nonhelical)
magnetic fields.

For an imposed magnetic field, on the other hand, the magnetic field is
always below the line $\EEM\propto\xiM^{-1}$, which corresponds to
the evolution of a fully helical magnetic field.
This is simply because magnetic helicity is no longer conserved in
that case; see the blue lines in the last panel of \Fig{pEkM2}.
The marked segregation between the blue and red lines in all panels of
\Fig{pEkM2} is an impressive and very visual demonstration of the tremendous
difference between the cases of an initial and an imposed magnetic field.
Only when the imposed magnetic field is weak enough are the two cases in
mutual in agreement with each other; compare the red (A) and blue (a)
lines in all panels of \Fig{pEkM2}.

Finally, we show in \Figs{pq}{pq2} the evolution of the instantaneous scaling
exponents $p(t)$ and $q(t)$ in a $pq$ diagram.
We see that in the case with an imposed magnetic field of moderate
strength in panel (a) the point $(p,q)$ appears to evolve along the line
$p=2(1-q)$ toward $(p,q)=(2,0)$.
This is indeed consistent with Fig.~\ref{pEkM2}, where $\EEM(t)$ is
seen to decay like $t^{-2}$ and $\xiM(t)$ is approximately flat.
For a stronger imposed field in panel (b), there seems to be an
evolution along $\beta=3$--$4$ toward $(p,q)\to(0,0)$, but
this is not consistent with Fig.~\ref{pEkM2}, where $\EEM(t)$ is
seen to decay like $t^{-4}$, while $\xiM(t)$ is still approximately flat.
Indeed, the points in \Fig{pq}(b) have a similar size, suggesting
that the evolution along the line $\beta=3$--$4$ is an intermediate
stage before later evolving toward $(p,q)\to(4,0)$, which is obviously
outside the plot range.

On the other hand, for an additional large-scale nonhelical
magnetic field, in addition to the small-scale helical one, the
evolution of the point $(p,q)$ always occurs along the $\beta=0$ line.
As time goes on, the point $(p,q)$ evolves further along the
$\beta=0$ line toward the left to smaller values of $p(t)$ and $q(t)$.
For the weak large-scale magnetic field of panel (c), the evolution
stalls near the point $(p,q)=(0.6,\,0.6)$.
Several intermediate points cluster along the line $p=0.58$, which was
identified in earlier work \cite{Brandenburg:2017rnt}, but this may
be coincidental.
Indeed, for the stronger large-scale magnetic field of panel (d), the
evolution continues toward the point $(p,q)=(0.2,\,0.2)$.
For sufficiently weak imposed magnetic fields, the cases of
imposed and initial magnetic fields again agree with each other;
compare \Fig{pq}(a) for run~(a) with \Fig{pq2}(A) for run~(A).

These investigations have demonstrated the dramatic difference
between imposed and initial large-scale magnetic fields.
When the imposed fields are weak, it only affects the evolution of the
small-scale helical magnetic field at later times once its field
strength approaches the value of the imposed field.
In the presence of a large-scale nonhelical magnetic field -- here one
with a $k^{-1}$ spectrum -- the inverse cascade is not suppressed.
Both for weak and strong magnetic fields, there is a spectral peak
moving from large to small wave numbers; see \Fig{pkt1152b}.
Also the evolution in the $pq$ diagram is along the $\beta=0$ line
in both cases; see \Fig{pq2}.

We emphasize again that the presence of an imposed field is pathological,
if the interest is to simulate an approximation to the case with a
large-scale magnetic field.
We have demonstrated this here with an irregular large-scale field with
a $k^{-1}$ spectrum.
Starting with an initially sinusoidal magnetic field is probably similar
in many ways, but this would introduce anisotropies, which have not yet
been studied in the context of decaying turbulence.

\section{Conclusions}
\label{secConcl}

We have discussed the viability of a homogeneous magnetic field after inflation.
Our work therefore extends the earlier work of one of the authors \cite{Mukohyama:2018obj},
which addressed only the stability of the magnetic field in the inflationary stage.
In this work, we have addressed the phenomenology of the primordial plasma
after inflation in the presence of a homogeneous magnetic field.
Our results apply to the early epochs of the Universe all the way from the time when the inequality in Eq. \eqref{eqn:tem-nonmim-coup-ignore} is satisfied until
matter radiation equality.

Our simulations have verified that, in the presence of an imposed magnetic
field, magnetic helicity is not conserved.
Moreover, and this was not previously known, our results demonstrate that
the decay of magnetic energy in the fluctuations is faster the stronger
the imposed magnetic field.
We have also compared the magnetic field evolution 
with an alternative way of simulating a cosmological
large-scale magnetic field, namely to treat it as a
statistically homogeneous field with a scale-invariant spectrum.
It is no longer the stealth magnetic field considered in the scenario of 
\cite{Mukohyama:2016npi}, but one that could emerge at the end of inflation.
Such a magnetic field can be either helical
\cite{Kahniashvili:2016bkp,Brandenburg:2018ptt}
or nonhelical \cite{Kahniashvili:2012vt}.
In these cases, magnetic helicity conservation is unaffected by the large-scale
magnetic field, and it decays just like without imposed magnetic
field and thus much more slowly than with a constant imposed magnetic field.

Conservation of magnetic helicity (and correspondingly its presence
until recombination) can have important observational consequences.
In particular, primordial magnetic helicity (as a manifestation of the
possible violation of parity in the early Universe) can leave traces  
in (i) the cosmic microwave background fluctuations, resulting in nonzero
temperature $B$-polarization, and $E$- and $B$-polarization cross correlations
(see \cite{Kahniashvili:2014dfa,Ballardini:2014jta} and references therein),
and (ii) the circular polarization of gravitational 
waves generated in the early Universe through helical hydrodynamical and
MHD turbulence (see \cite{Pol:2019yex} and references therein).

As for the backreaction of small-scale fields to the background magnetic
field at large scale, \textit{a priori} there could be three possibilities: (i)
small-scale fields inverse cascade and deplete the background magnetic
field at large scale; (ii) small-scale fields inverse cascade and enhance
the background magnetic field at large scale; or (iii) small-scale
fields do not affect the background magnetic field at large scale.
The result of the present paper suggests that (iii) is the case.
Therefore, a homogeneous magnetic field, if generated during inflation,
should persist [i.e., simply decay as $\propto 1/a^2$, as assumed in
the derivation of (\ref{eqn:Btoday})] under the influence of small-scale
fields and could be the origin of the large-scale magnetic field in the
Universe today.
Depending on the strength of the background magnetic field, however,
the small-scale magnetic field can be significantly suppressed.
The low power at small scales (see the blue lines in \Fig{pEkM2})
means that the homogeneous background may dominate the magnetic fields
in the Universe  not only at large scales but also at small scales.
This implies that, depending on its strength, the background magnetic
field may be responsible not only for the blazar observations, but also
for the seeds of MHD processes at astrophysical scales
such as galactic dynamo.

\vspace{5mm}

\acknowledgements

We thank Arthur Kosowsky and Alexander Tevzadze for useful discussions.
T.K. and S.M. (CMU) are especially grateful for the hospitality provided by the
University of Geneva, where initial discussions on this project were made.
Support through the NSF Astrophysics and Astronomy Grant Program
(Grants No. 1615940 and No. 1615100), and the Shota Rustaveli NSF
(Georgia) (Grants No. FR/18-1462 and No. FR/19-8306) are gratefully acknowledged.
The work of S.M. (YITP) was supported by Japan Society for the Promotion
of Science Grants-in-Aid for Scientific Research No.~17H02890,
No.~17H06359, and by World Premier International Research Center
Initiative, MEXT, Japan. 
R.D. is supported by the Swiss National Science Foundation.
We acknowledge the allocation of computing resources provided by the
Swedish National Allocations Committee at the Center for Parallel
Computers at the Royal Institute of Technology in Stockholm.

\appendix
\section{Magnetic helicity evolution in the presence of a homogeneous magnetic field}
\label{app1}

The purpose of this appendix is to recall why magnetic helicity is not a
conserved quantity in the presence of a homogeneous or imposed magnetic
field, $\BB_0$.
We write $\BB=\BB_0+\mathbf{b}$, where $\mathbf{b}=\nab\times\mathbf{a}$,
with $\mathbf{a}$ being the vector potential of $\mathbf{b}$.
Equation~\eqref{dAdt} can then be written in terms of $\mathbf{a}$ as
\begin{equation}
\frac{\partial\mathbf{a}}{\partial t}=\mathbf{u}\times\mathbf{B}_0
+\mathbf{u}\times\mathbf{b}-\eta\mathbf{j}+\EEE_0.
\label{dadt}
\end{equation}
We consider times $t>t_\ast$ when $\EEE_0=\bm{0}$.
Making use of the periodic boundary conditions for $\mathbf{a}$,
the evolution equation for $\bra{\mathbf{a}\cdot\mathbf{b}}$ is then
\begin{equation}
\frac{d}{dt}\bra{\mathbf{a}\cdot\mathbf{b}}=
2\bra{(\mathbf{u}\times\mathbf{B}_0)\cdot\mathbf{b}}
-2\eta\bra{\mathbf{j}\cdot\mathbf{b}}.
\label{dabdt}
\end{equation}
Evidently, the first term on the right-hand side of Eq.~\eqref{dabdt}
breaks magnetic helicity conservation in the limit $\eta\to0$, because
\begin{equation}
\bra{(\mathbf{u}\times\mathbf{B}_0)\cdot\mathbf{b}}
=-\bra{\mathbf{u}\times\mathbf{b}}\cdot\mathbf{B}_0
=-\alpha\mathbf{B}_0^2\neq0
\label{dalpdt}
\end{equation}
for a helical magnetic field, where $\alpha\neq0$.
We recall an important result from mean-field electrodynamics \cite{Mof78,KR80}
\begin{equation}
\bra{\mathbf{u}\times\mathbf{b}}_i=\alpha_{ij} \bra{B_j}+\beta_{ijk} \partial\bra{B_j}/\partial x_k
+\cdots,
\end{equation}
which reduces to $\bra{\mathbf{u}\times\mathbf{b}}=\alpha\mathbf{B}_0$
for periodic boundary conditions.
Here, the ellipsis refers to higher order terms.
The nonconservation of magnetic helicity is not an artifact of having
adopted periodic boundary conditions, because they are just a tool
for us to compute averages over infinitely large length scales.

\end{document}